\documentclass[]{aa}  
\usepackage{amsmath}
\usepackage{amssymb}
\usepackage{natbib}
\usepackage{graphicx}
\usepackage{txfonts}
\usepackage[english]{babel}
\usepackage{units}
\begin{document}
\title{Radiative transfer in very optically thick circumstellar disks}

\author{
	M. Min\inst{1}
		\and
	C.~P. Dullemond\inst{2}
		\and
	C. Dominik\inst{1}
		\and
	A. de Koter\inst{1,3}
		\and
	J.~W. Hovenier\inst{1}
}

\offprints{M. Min, \email{mmin@science.uva.nl}}

\institute{
Astronomical institute Anton Pannekoek, University of Amsterdam,
Kruislaan 403, 1098 SJ  Amsterdam, The Netherlands
	\and
Max-Planck-Institut f\"ur Astronomie Heidelberg, K\"onigstuhl 17, D69117 Heidelberg, 
Germany
	\and
Astronomical institute Utrecht, University of Utrecht, P.O. Box 80000, NL-3508 TA Utrecht, The Netherlands
}

   \date{Received September 15, 1996; accepted February 11, 2009}

 
  \abstract
   {}
   {In this paper we present two efficient implementations of the diffusion approximation to be employed in Monte Carlo computations of radiative transfer in dusty media of massive circumstellar disks. The aim is to improve the accuracy of the computed temperature structure and to decrease the computation time. The accuracy, efficiency and applicability of the methods in various corners of parameter space are investigated. The effects of using these methods on the vertical structure of the circumstellar disk as obtained from hydrostatic equilibrium computations are also addressed.}
   {Two methods are presented. First, an energy diffusion approximation is used to improve the accuracy of the temperature structure in highly obscured regions of the disk, where photon counts are low. Second, a modified random walk approximation is employed to decrease the computation time. This modified random walk ensures that the photons that end up in the high-density regions can quickly escape to the lower density regions, while the energy deposited by these photons in the disk is still computed accurately. A new radiative transfer code, MCMax, is presented in which both these diffusion approximations are implemented. These can be used simultaneously to increase both computational speed and decrease statistical noise.}
   {We conclude that the diffusion approximations allow for fast and accurate computations of the temperature structure, vertical disk structure and observables of very optically thick circumstellar disks.}
   {}


   \maketitle
%

\section{Introduction}

Protoplanetary disks are the sites of planet formation and thus are the ideal environments for studying how planetary systems are formed. With the current development of new observational techniques and the increasing sensitivity of modern telescopes, the call for accurate, fast and flexible methods of interpreting these observations is increasing. Analysis tools are needed that are fast and, at the same time, that do justice to the rich detail of current observations.

An important step in translating predictions from disk properties into comparison with observations lies in solving the radiative transfer in protoplanetary disks subject to conservation of energy and thermal equilibrium. The opacity and energy balance in protoplanetary disks are dominated by their dust grains, even though they only contribute about one percent of the total disk mass. Unfortunately, determining the radiation transport is far from a trivial task, mainly because of the wide range of optical depths that have to be traced in the disk. On the one hand, the optically thin upper layers of the disk are important since they are the regions where solid-state resonances are formed that give information on the size distribution and composition of the dust grains in these layers. On the other hand, the cold midplane region of the disk is important since it contains most of the mass of the disk and thereby determines the vertical structure. Radial optical depths through this midplane can easily reach values of $10^5$-$10^6$ in the visible part of the spectrum.

The required flexibility for radiative transfer can be obtained by using a Monte Carlo method \citep[see e.g.][]{2001ApJ...554..615B, 2003A&A...399..703N, 2006A&A...459..797P}. Its advantage is that it is accurate at all optical depths and puts virtually no restrictions on the spatial distribution of matter or on the optical properties of the dust grains. The disadvantage is that the method becomes increasingly slow for high optical depths since every interaction of a photon package has to be considered separately. Also, in the deep layers of the disk that are well-shielded from the stellar radiation, photon counts may be low, causing large errors on the derived temperature structures in these regions.
Accurate values for the temperatures in the midplane regions of protoplanetary disks are crucial for computing self-consistent vertical density distributions \citep{2004A&A...417..159D, 2007prpl.conf..555D}.

In this paper we discuss solutions to these problems in terms of two different formulations of the diffusion approximation for high optical depth regions. The two different formulations serve different purposes and can be used together. We present a partial diffusion approximation (PDA) which can be used to decrease statistical noise on the temperature structure and a modified random walk (MRW) procedure which is useful for decreasing computation time significantly. Both used together provide a fast and accurate radiative transfer scheme that can be used for a variety of applications. We focus on passive protoplanetary disks, i.e. where the radiation from the central star dominates, and consider only the radiative transfer through the dust component. In Sect.~\ref{sec:computational approach} we will outline the general radiative transfer method employed along with the implementations of the PDA and MRW. Then, in Sect.~\ref{sec:results} we will check the accuracy of the approximations and discuss the resulting self-consistent vertical density structures. The results are put together into recommendations for the use of the diffusion approximation in various cases in Sect.~\ref{sec:recommendations}. Finally, in Sect.~\ref{sec:conclusions} we summarize our findings and draw our conclusions.

\section{Computational approach}
\label{sec:computational approach}

\subsection{Monte Carlo radiative transfer}

A flexible way of computing the transfer of radiation in an inhomogeneous, complex medium is by means of a Monte Carlo method. In this method photon packages emerging from the central star are traced through the disk, allowing them to undergo scattering, absorption and reemission events caused by the dust they encounter on their path. The Monte Carlo method for radiative transfer in dusty media has become particularly efficient since the ideas of continuous absorption and reemission have been proposed and worked out by \citet{2001ApJ...554..615B}. The method is fully described in their paper, and we will not repeat it here. 

Two major difficulties of Monte Carlo radiative transfer are:
\begin{enumerate}
\item The statistical noise in regions shielded from radiation.
\item The fact that computation time increases with roughly the square of the optical depth.
\end{enumerate}

At first glance, the first difficulty might seem not so relevant since in radiative transfer in protoplanetary disks the noise mainly appears in regions which are optically well shielded, i.e. mainly the midplane of the disk, and the midplane contributes only little to the observed infrared spectrum. However, for images, visibilities and millimeter observations it can be of importance to have a reliable estimate of the midplane temperature. In addition, the midplane temperature is crucial for hydrostatic equilibrium computations of the vertical structure of the disk. For this reason, computations of the vertical structure of protoplanetary disks have, up to now, relied on more simplified methods that do not have the problem of noise. We will in the next section outline a method to circumvent this problem, and obtain a reliable approximation of the temperature in regions with low photon counts.

\subsection{Diffusion approximation for optically thick regions}
\label{sec:diffusion}

\subsubsection{Decreasing photon noise: Partial Diffusion Approximation (PDA)}
\label{sec:Radiative diffusion}

In regions in the disk where the mean free path of the photons is much smaller than the length scale over which density, $\rho$, and temperature, $T$, change, the energy transport is by radiative diffusion. This is described by the radiative diffusion equation (see \cite{2000A&A...359..780W} who provide an extension of the classical one dimensional result by \cite{1924MNRAS..84..525R} to three dimensions)
\begin{equation}
\label{eq:diffusion}
\nabla\cdot\left(D\nabla E\right)=\frac{1}{c}\frac{\partial E}{\partial t},
\end{equation}
where $E$ is the local energy density, $c$ is the speed of light, and $D$ is the diffusion coefficient which in the Rosseland approximation is given by
\begin{equation}
\label{eq:RossDiff}
D_R=\frac{1}{3}\lambda_R=\frac{1}{3\rho\bar{\chi}_R}
\end{equation}
where $\lambda_R=(\rho\bar{\chi}_R)^{-1}$ is the Rosseland mean free path, and $\bar{\chi}_R$ is the Rosseland mean opacity given by
\begin{equation}
\frac{1}{\bar{\chi}_R}=\frac{\displaystyle{\int_0^\infty \chi_\nu^{-1}\quad\frac{\partial B_\nu}{\partial T}d\nu}}{ \displaystyle{\int_0^\infty\frac{\partial B_\nu}{\partial T}d\nu}}.
\end{equation}
Here $\chi_\nu$ is the wavelength dependent mass extinction coefficient, $B_\nu$ is the Planck function, and $\nu$ is the frequency of radiation.
The Rosseland diffusion coefficient is an approximate measure. In case scattering cannot be neglected, the diffusion coefficient has to be adjusted. The expression for the diffusion coefficient we use is given in section~\ref{sec:diffcoeff}.

We can use the diffusion equation to improve the accuracy of the temperatures in regions of high optical depth. The way we implement this is that after the Monte Carlo procedure we solve the time independent version of Eq.~(\ref{eq:diffusion}), i.e. we put $\partial E/\partial t=0$ in regions of low photon counts. In this case we get the diffusion approximation for the temperature in the cell
\begin{equation}
\label{eq:T diffusion}
\nabla\cdot\left(D\nabla T^4\right)=0.
\end{equation}
Eq.~(\ref{eq:T diffusion}) results in a system of linear equations which can be solved if we set appropriate boundary conditions. Since we wish to solve the diffusion equation only in a limited region of the disk (the regions around the midplane), the boundary conditions are simply set by the temperatures as determined by the Monte Carlo procedure at the edge of the region with low photon counts. We refer to this method as the Partial Diffusion Approximation (PDA). The accuracy of the results of this approximation will be discussed in detail later in this paper. Note that the PDA is only used \emph{after} the full Monte Carlo run has already finished. Thus observables that are directly obtained from the escaping Monte Carlo photon packages are not influenced by it. The main use of the PDA is to obtain a reliable temperature structure for, for example, iteratively solving the vertical hydrostatic density distribution.

The PDA assumes that the probability that a photon escapes from the optically thick region without interactions is zero. In reality though, a photon emitted at a very long wavelength encounters a very low opacity, and thus the disk can become optically thin for these photons. This will cause these regions to cool more efficiently than predicted by the PDA. Thus, in general the PDA is expected to overestimate the temperature slightly. We will return to this point when we discuss the accuracy of the PDA.

\subsubsection{Increasing computational speed: Modified Random Walk (MRW)}
\label{sec:RW}

Monte Carlo radiative transfer computations can be very slow when implemented in a straightforward manner, especially for geometrical configurations where a photon package has a large number of interactions with the dust before it leaves the system. For protoplanetary disks this occurs especially when a single photon package 'gets lost' in the optically thick regions of the disk, i.e. the disk midplane. It can then take a very high number of interactions before the photon package finally leaves the system. Although only very few photons get to these regions, in some configurations of rather massive protoplanetary disks these photons completely dominate the computation time. We can try to overcome this problem by letting photons make multiple interaction steps in a single computation. This can be done by using the radiative diffusion equation as outlined below.

In the regions of radiative diffusion the photons travel by a random walk. This random walk is basically a solution to the radiative diffusion equation (Eq.~\ref{eq:diffusion}).
In the Monte Carlo procedure we wish to know the distance a photon travels before leaving a given region in the disk. 
For the Monte Carlo procedure, space is subdivided in regions of constant density and temperature and we can apply the method of \citet{1984JCoPh..54..508F}. The solution to Eq.~(\ref{eq:diffusion}) in the case of an infinite homogeneous medium is simply
\begin{equation}
\label{eq:RW}
\psi(\mathbf{r},t)=\frac{1}{(4\pi D\,ct)^{3/2}}\exp\left(-\frac{r^2}{4D\,ct}\right),
\end{equation}
where $\psi(\mathbf{r},t)$ is the fraction of the energy that has diffused to position $\mathbf{r}$ in a time $t$, and $r=|\mathbf{r}|$.

Eq.~(\ref{eq:RW}) is only valid for an infinite, homogeneous medium while the cells in the disk are of finite size. The way to solve this is outlined by \citet{1984JCoPh..54..508F} and consists of setting an absorbing boundary condition to the diffusion equation at $r=R_0$, with $R_0$ being the radial boundary of the homogeneous region. This ensures that the time it takes for the photon to reach the boundary of the region is the time when the photon crosses the boundary \emph{for the first time}. The solution to this is a Modified Random Walk (MRW) and is given by
\begin{equation}
\label{eq:modified RW}
\psi(\mathbf{r},t)=\frac{1}{R_0^2}\sum_{n=1}^{\infty} \frac{n}{r}~\exp\left\{-\left(\frac{\pi n}{R_0}\right)^2 D\,ct\right\} \sin\left(\frac{n\pi r}{R_0}\right).
\end{equation}
To determine the likelihood that the photon is still inside a sphere with radius $R_0$ after a time $t$ we have to integrate $\psi(\mathbf{r},t)$ from $0$ to $R_0$
\begin{equation}
\label{eq:probability RW}
P(t)=4\pi\int_0^{R_0} r^2 \psi(\mathbf{r},t) dr=2\sum_{n=1}^{\infty} (-1)^{n+1}y^{n^2}
\end{equation}
with
\begin{equation}
y=\exp\left\{-D\,ct\left(\frac{\pi}{R_0}\right)^2\right\}.
\end{equation}
In the Monte Carlo procedure we now draw a random number, $P_0$, between $0$ and $1$ and solve $P(t)=P_0$ for $t$. This now gives us the time the photon has traveled before reaching the edge of the sphere. The energy that is absorbed by the dust grains per unit mass during this random walk to the edge of the sphere is equal to 
\begin{equation}
E=E_\gamma ct\rho\bar{\kappa},
\end{equation}
where $E_\gamma$ is the energy of the photon package, and $\bar{\kappa}$ is the average mass absorption coefficient encountered by the photons
\begin{equation}
\bar{\kappa}=\frac{\displaystyle{\int_0^\infty \frac{\kappa_\nu}{\chi_\nu}~\eta_\nu~d\nu}}{ \displaystyle{\int_0^\infty \chi_\nu^{-1}\eta_\nu~d\nu}}.
\end{equation}
where $\eta_\nu$ is the emission coefficient, and $\kappa_\nu$ is the wavelength dependent mass absorption coefficient. The emission coefficient $\eta_\nu$ is proportional to the probability that a photon leaves an interaction event (scattering or re-emission) with frequency $\nu$. An iterative way to obtain $\eta_\nu$ is given in section~\ref{sec:diffcoeff}. The above equation averages the frequency-dependent mass absorption coefficient weighted with the average distance traveled at that particular frequency.
This procedure can then be repeated until the photon is close enough to the cell's edge where the regular Monte Carlo method can take over.

Since the MRW is only valid in optically very thick regions, an important aspect of the above method is a criterion to start the MRW procedure. A good criterion for this is that the smallest distance to the cell's edge is larger than a few times the Rosseland mean free path \citep{1984JCoPh..54..508F}, i.e.
\begin{equation}
\label{eq:criterion}
d_{\mathrm{min}}>\frac{\gamma}{\rho\bar{\chi}_R}.
\end{equation}
The parameter $\gamma$ can be adjusted for speed or accuracy. Values of $\gamma$ close to unity ensure fast radiative transfer but with relatively large errors, while higher values of $\gamma$ ensure more accurate results but at the cost of a slightly increased runtime. In addition to this criterion we also require the photon package to have experienced several tens of interactions in a single cell before the MRW is employed. Note that there is always a finite chance that a photon gets emitted at a wavelength where the optical depth is low and the photon can escape the cell without interactions. The criterion set in Eq.~(\ref{eq:criterion}) has to be chosen such that the fraction of such photons is small.
We study the effects of different values of $\gamma$ in section~\ref{sec:random walk test}.

\subsubsection{The Diffusion Coefficient}
\label{sec:diffcoeff}

In this section we derive the diffusion coefficient used in the modified random walk procedure and the partial diffusion approximation. We start with the diffusion coefficient as given by \citet{1984JCoPh..54..508F} for the modified random walk
\begin{equation}
\label{eq:diffusion coeff}
D=\frac{\left<d^2\right>}{6\left<d\right>}
\end{equation}
where $d$ is the distance a photon package travels, $\left<d\right>$ is the mean free path, and $\left<d^2\right>$ is the mean of the pathlength squared. For monochromatic radiative transport at frequency $\nu$ we have that $\left<d\right>=d=(\rho\chi_\nu)^{-1}$ and $\left<d^2\right>=d^2=2(\rho\chi_\nu)^{-2}$. 

Since we have absorption and reemission in addition to scattering, we have to consider a broad range of wavelengths. Therefore, we have to average the mean free path over all frequencies using a proper weighting function. The proper weighting function is the energy probability distribution leaving an interaction (absorption/reemission or scattering) event which is proportional to the emission coefficient $\eta_\nu$. This emission coefficient obeys the following integral equation 
\begin{equation}
\label{eq:eta nu iter}
\eta_\nu=\kappa_\nu\frac{\partial B_\nu}{\partial T}~\frac{\displaystyle{\int_0^\infty \eta_\nu\,\frac{\kappa_\nu}{\chi_\nu}\,d\nu}}{\displaystyle{\int_0^\infty \eta_\nu\,d\nu}}~+~\frac{\sigma_\nu}{\chi_\nu}\eta_\nu,
\end{equation}
where $\sigma_\nu$ is the wavelength dependent mass scattering coefficient. Eq.~(\ref{eq:eta nu iter}) can be solved iteratively. When an guess of $\eta_\nu$ is available, an improved estimate can be obtained by substituting this guess into the right hand side of Eq.~(\ref{eq:eta nu iter}). By repeating this procedure several times the solution is easily obtained.
A good initial guess in most cases is to ignore scattering and take $\eta_\nu=\kappa_\nu\partial B_\nu/\partial T$.

Usually the procedure described above converges in a few iterations. When $\eta_\nu$ is computed $\left<d\right>$ and $\left<d^2\right>$ are given by
\begin{equation}
\label{eq:avd}
\left<d\right>=\frac{\displaystyle{\int_0^\infty \frac{\eta_\nu}{\rho\chi_\nu}\,d\nu}}{\displaystyle{\int_0^\infty \eta_\nu\,d\nu}},
\end{equation}
and
\begin{equation}
\label{eq:avd2}
\left<d^2\right>=2~\frac{\displaystyle{\int_0^\infty \frac{\eta_\nu}{\rho^2\chi_\nu^2}\,d\nu}}{\displaystyle{\int_0^\infty \eta_\nu\,d\nu}}.
\end{equation}
Note that when there is no scattering, i.e. $\chi_\nu=\kappa_\nu$, Eqs.~(\ref{eq:diffusion coeff}-\ref{eq:avd2}) reduce to the Rosseland diffusion coefficient given by Eq.~(\ref{eq:RossDiff}).

The formulae above hold for the case of isotropic scattering, which we consider in this paper. However, simple corrections can be made for anisotropic scattering. 
As shown by \citet{1978usaf.book.....I}
in the diffusion domain similar results are obtained when $\sigma_\nu$ and the asymmetry parameter $g$ are both replaced by $\sigma_\nu'$ and $g'$ given by 
\begin{equation}
\frac{\sigma_\nu}{\sigma_\nu'}=\frac{1-g'}{1-g}.
\end{equation}
The simplest form is by taking the solution for isotropic scattering, i.e. $g'=0$, with $\sigma_\nu'=(1-g)\sigma_\nu$.

\subsection{Description of the code: MCMax}

The radiative transfer code MCMax is based on the Monte Carlo method outlined by \citet{2001ApJ...554..615B} to solve the temperature structure in the disk. The observables are obtained by analytic integration of the computed source function in order to avoid noise in the observables. The partial diffusion approximation as well as the modified random walk as discussed above are implemented as options. The code solves the radiative transfer in 3 dimensions but it is assumed that the geometry is axisymmetric. Sophisticated regridding algorithms make sure that the optical depth to the stellar and local radiation field is properly sampled so that the temperature structures of extreme density distributions can be solved accurately.

The scattering of radiation by the dust grains is treated in a complete sense. This means that the full angle-dependent Mueller matrix is used for each scattering event. In this way, also polarization maps of the disks can be obtained. A comparison of the results for anisotropic scattering with other codes will be presented in \citet{PinteBenchmark}. In this study we assume isotropic scattering.

In addition to solving the radiative transfer equations subject to the constraint of radiative equilibrium, the MCMax code can also be used to solve the vertical scale height of the disk self consistently by iterating the radiative transfer computations and subsequently solving vertical hydrostatic equilibrium equations as outlined in section \ref{sec:vertical distr}.

\subsection{Disk setup}
\label{sec:disk setup}

For the density structure of the disk we follow partly the benchmark setup as defined by \citet{2004A&A...417..793P}. However, we modify the setup to more massive disks with higher optical depths. We deviate from the \citet{2004A&A...417..793P} setup on three important points. 
\begin{enumerate}
\item The surface density of the disk, $\Sigma$, as chosen by \citet{2004A&A...417..793P} increases with the distance to the central star $R$ as $\Sigma\propto R^{0.125}$. This implies that the vertical optical depth through the disk increases with distance, putting the most optically thick regions far away from the star. We consider a radial dependence of the surface density which is a power law with $\Sigma\propto R^{-1}$ \citep[which is closer to the theoretically predicted value, see][]{2007prpl.conf..555D}. 
\item The outer disk radius in the \citet{2004A&A...417..793P} setup is chosen very large, $R_\mathrm{out}=1000\,$AU, while protoplanetary disks are thought to have smaller sizes. We therefore adopt a value of $R_\mathrm{out}=200\,$AU.
\item For the vertical scale height of the disk we choose a parameterization which is close to the case of vertical hydrostatic equilibrium. In section~\ref{sec:vertical structure} we will show the results for the iterated solution of the vertical density distribution.
\end{enumerate}

For the scale height of the disk we also take a powerlaw with distance, $h(R)\propto R^p$, where $p$ is determined to be roughly consistent with that of a disk in vertical hydrostatic equilibrium. The density structure we take is for $R_\mathrm{in}<R<R_\mathrm{out}$ given by
\begin{eqnarray}
\label{eq:density}
\rho(R,z) & = & \frac{M_\mathrm{dust}(1\,\mathrm{AU})^p} {2\pi\sqrt{\pi}~h_1 \left(R_\mathrm{out}-R_\mathrm{in}\right)} \, R^{-1-p}~e^{-\left(z/h(R)\right)^2}, \\
h(R)&=&h_1 \left(\frac{R}{1\,\mathrm{AU}}\right)^p.
\end{eqnarray}
Here $\rho(R,z)$ is the dust density in the disk, $M_\mathrm{dust}$ is the total dust mass, $R$ is the distance to the center of the star (in cylindrical coordinates), $z$ is the height above the midplane, $p$ is the power of the scaleheight powerlaw, $h_1$ is the scaleheight of the disk at $1\,$AU, and $R_\mathrm{in}$ and $R_\mathrm{out}$ are the inner and outer radius of the disk. For the values of the parameters see Table~\ref{tab:parameters}. The radius of the inner edge of the disk is chosen to correspond roughly to the dust condensation temperature.

By integrating Eq.~(\ref{eq:density}) through the midplane ($z=0$) from $R=R_\mathrm{in}$ to $R_\mathrm{out}$ we can compute the radial optical depth
\begin{equation}
\label{eq:radial tau}
\tau_{\mathrm{radial},\nu} = \int_{R_\mathrm{in}}^{R_\mathrm{out}} \chi_\nu~\rho(R,0) dR=\frac{M_\mathrm{dust}\left(R_\mathrm{in}^{-p}-R_\mathrm{out}^{-p}\right)(1\,\mathrm{AU})^p}{2\pi\sqrt{\pi}~h_1 p \left(R_\mathrm{out}-R_\mathrm{in}\right)}~\chi_\nu.
\end{equation} 
This corresponds, for the parameters of the disks we have chosen, to an optical depth at 550\,nm of $\tau_{\mathrm{radial},550\,\mathrm{nm}}=3.5\cdot 10^6\,(3.5\cdot 10^3)$ for the model with $M_\mathrm{dust}=10^{-3}\,(10^{-6})\,M_{\sun}$.

We take a very simplistic description of the grain properties. The dust grains in the disk are homogeneous spherical particles with a radius of $0.12\,\mu$m composed of so-called 'astronomical silicate' \citep{1984ApJ...285...89D}. The scattering by the grains is assumed to be isotropic. We would like to note here that although the grains have in principle a non-isotropic phase function, the errors on the temperature structure and emerging spectra made by assuming isotropic scattering are generally small. So although both codes used for the calculations can in principle handle anisotropic scattering, we choose to use isotropic scattering for the computations in this paper.

\begin{table}[!t]
\caption{The parameters for the model setup (as used in Eq.~\ref{eq:density}).}
\begin{center}
\begin{tabular}{lll}
\hline
Symbol & Meaning   & Value\\
\hline
$M_\star$			& Stellar mass					& $2.5\,M_{\sun}$ \\
$R_\star$			& Stellar radius					& $2\,R_{\sun}$ \\
$T_\star$			& Stellar effective temperature		& $10000\,$K \\
$R_\mathrm{in}$	& Inner disk radius				& $0.75\,$AU \\
$R_\mathrm{out}$	& Outer disk radius				& $200\,$AU \\
$h_1$			& Scale height at $1\,$AU			& $0.03\,$AU \\
$p$				& Power law for the scale height		& $1.3$ \\
$a$				& Grain radius					& $0.12\,\mu$m \\
$\rho_g$			& Grain density					& $3.6\,$g/cm$^3$ \\
$M_\mathrm{dust}$	& Dust mass in the disk			& $10^{-6},\,10^{-3}\,M_{\sun}$ \\
\hline
\end{tabular}
\end{center}
\label{tab:parameters}
\end{table}

\subsection{Vertical structure}
\label{sec:vertical distr}

At first we will discuss the results using the above parameterized version of the density structure. In section \ref{sec:vertical structure} we solve the vertical structure using hydrostatic equilibrium computations and iterate on the vertical structure until it sufficiently converges. For these computations we kept the surface density fixed at $\Sigma\propto R^{-1}$. The vertical density profile is then set by the balance between pressure and gravity and is given by the solution of the following differential equation \citep[see e.g.][]{2002A&A...395..853D}.
\begin{equation}
\label{eq:balance}
\frac{dP(R,z)}{dz}=-\rho(R,z)~\frac{GM_\star}{R^3}~z,
\end{equation}
where $G$ is the gravitational constant, and the pressure $P=k_B\rho T/(\mu m_u)$ with $k_B$ the Boltzmann constant, the mean molecular weight $\mu=2.3$ (for a H$_2$, He mixture) and $m_u$ is the proton mass. The procedure is now to first compute the temperature structure for an initial guess of the density distribution. These temperatures and their derivatives can then be used to solve Eq.~(\ref{eq:balance}), and this can be iterated several times until the convergence criterion is met.

In order to judge if the vertical structure is sufficiently converged, a convergence criterion is needed. For this we used the temperature structure of the disk. The error in the temperature structure can be computed using the statistics from the Monte-Carlo procedure. This allows us to compute the difference between two iterations as compared to the statistical errors. When the average difference between the temperature structures of two consecutive iterations is less than $3\sigma$ we conclude that the model is converged.

\section{Results}
\label{sec:results}

\begin{figure*}[!t]
\resizebox{\hsize}{!}{\includegraphics{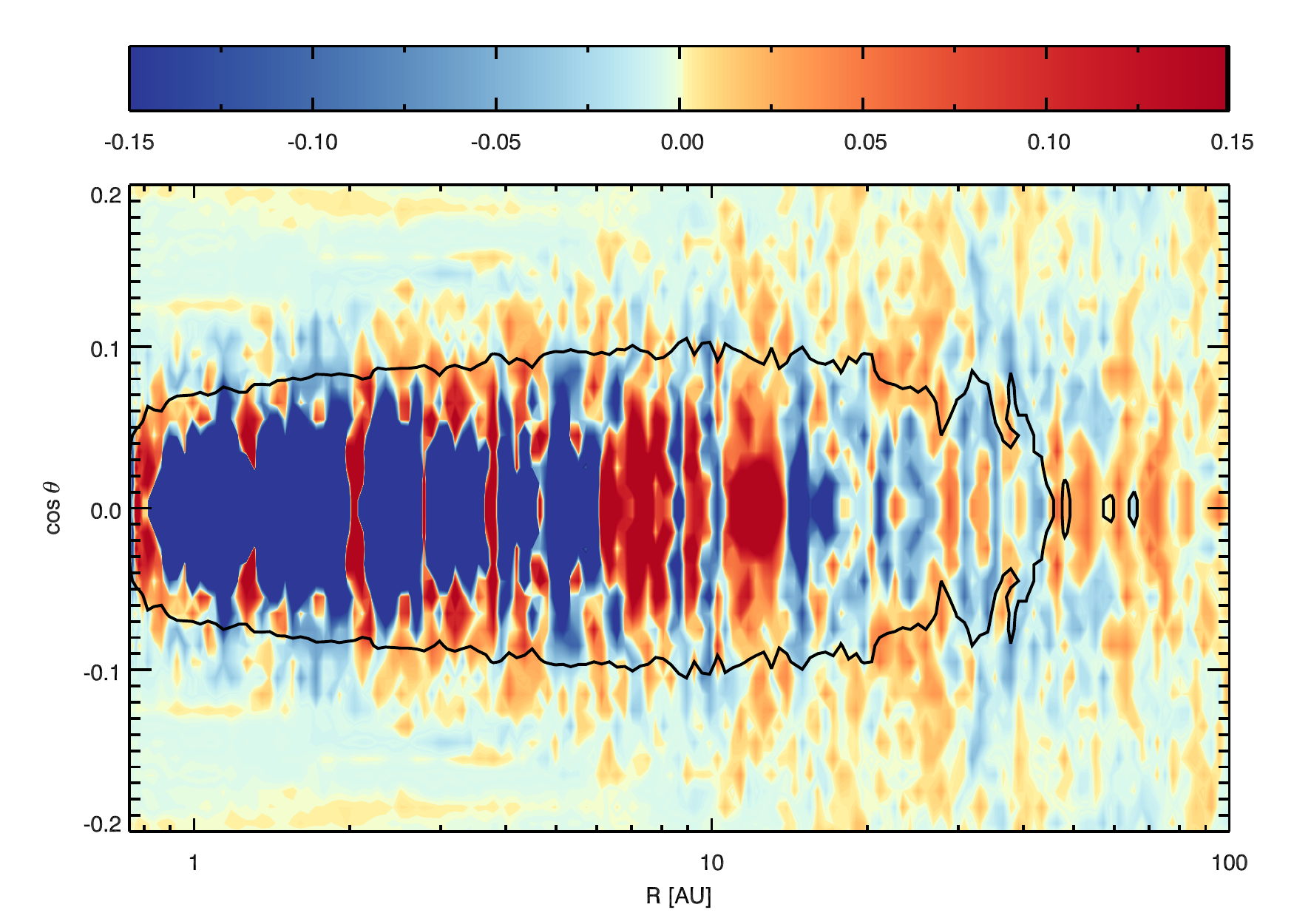}\includegraphics{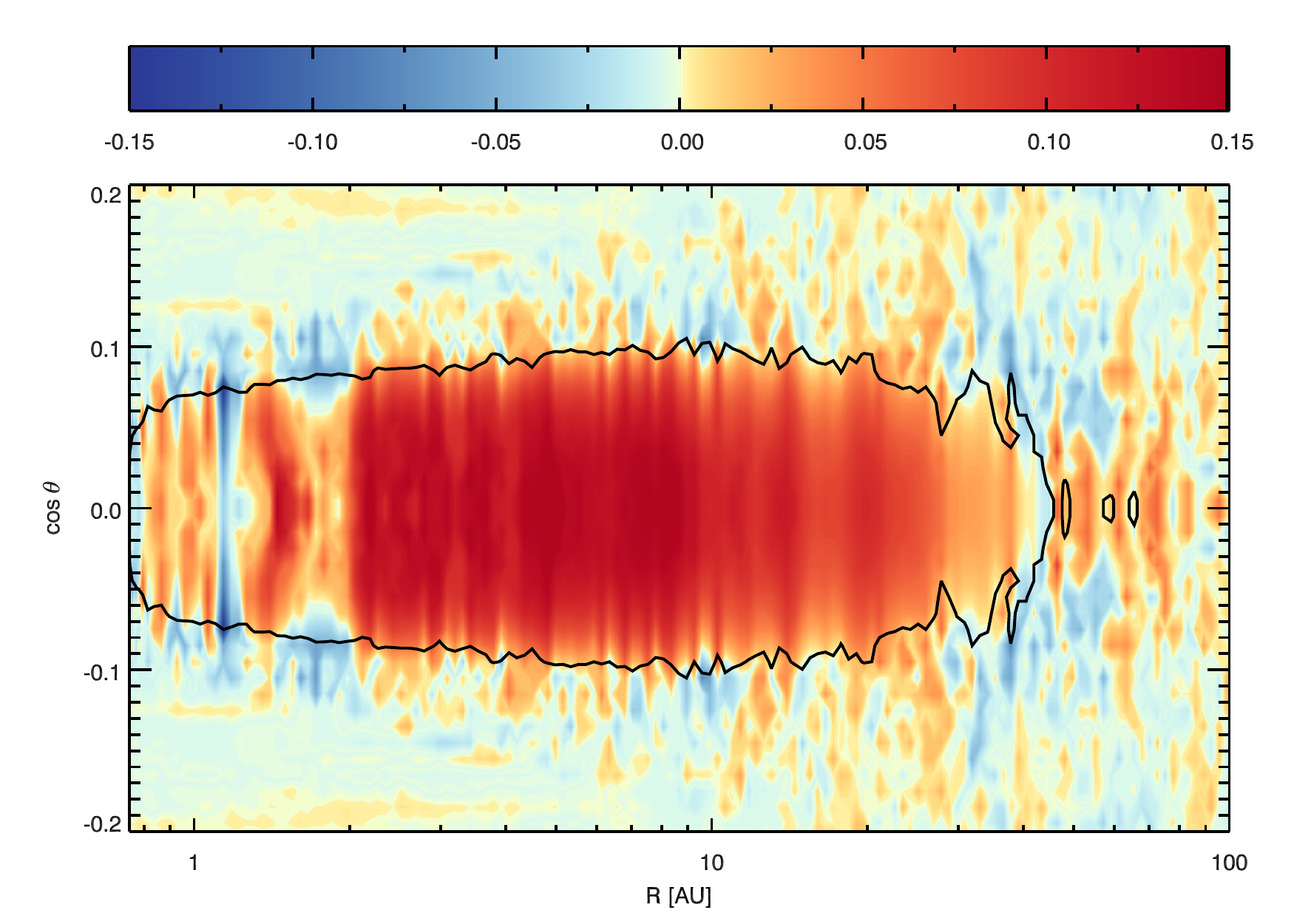}}
\caption{The relative error in the temperature computed by the model using $10^5$ photon packages as compared to the reference model using $10^8$ photon packages as a function of position in the disk. The left panel shows the model without the radiative diffusion approximation, the right panel with the radiative diffusion approximation. The black contour line encloses the region where the number of photon packages entering a cell is below $30$, which corresponds to an error in the derived temperature of $\sim5\%$. The dust mass in these models is $10^{-3}\,M_{\sun}$.}
\label{fig:compare diffusion}
\end{figure*}

\subsection{Results with partial diffusion approximation}
\label{sec:diffusion test}

In this section we will compare the results with and without the partial diffusion approximation (PDA) as outlined in Section~\ref{sec:diffusion}. We will compare the results of the runs with $10^5$ photon packages with and without radiative diffusion with those obtained using $10^8$ photon packages without the partial diffusion approximation or modified random walk procedure. We will refer to the latter model as the reference model. We have used the PDA everywhere where in the $10^5$ photon model the number of photons visiting a cell is below 30 or the relative error in the temperature as derived from the photon statistics is higher than 5\%. The relative difference in the computed temperature structure compared to the run with $10^8$ photon packages is shown in Fig.~\ref{fig:compare diffusion}. In this figure negative errors correspond to the case where the temperature using the PDA is lower than that of the reference model, while positive errors represent higher temperatures. It is clear that the PDA improves the agreement with the high photon run considerably, although there are still regions with relatively large errors. These errors are partly caused by errors in the temperature at the boundary of the region where the PDA is used. Most of the diffusion occurs vertically. Since the cells at the boundary are used to set the boundary conditions of the PDA, an error in the temperature in these cells propagates all the way to the midplane, causing the vertical lines of large errors seen in Fig.~\ref{fig:compare diffusion}.

The largest errors in the derived temperatures without the use of the PDA are $100\%$ (which lies outside the limits of errors shown in Fig.~\ref{fig:compare diffusion}). This is in the regions of the disk where there are no photon packages in the $10^5$ photon run.
By using the PDA we can reduce this error to $18\%$.

\begin{figure}[!t]
\resizebox{\hsize}{!}{\includegraphics{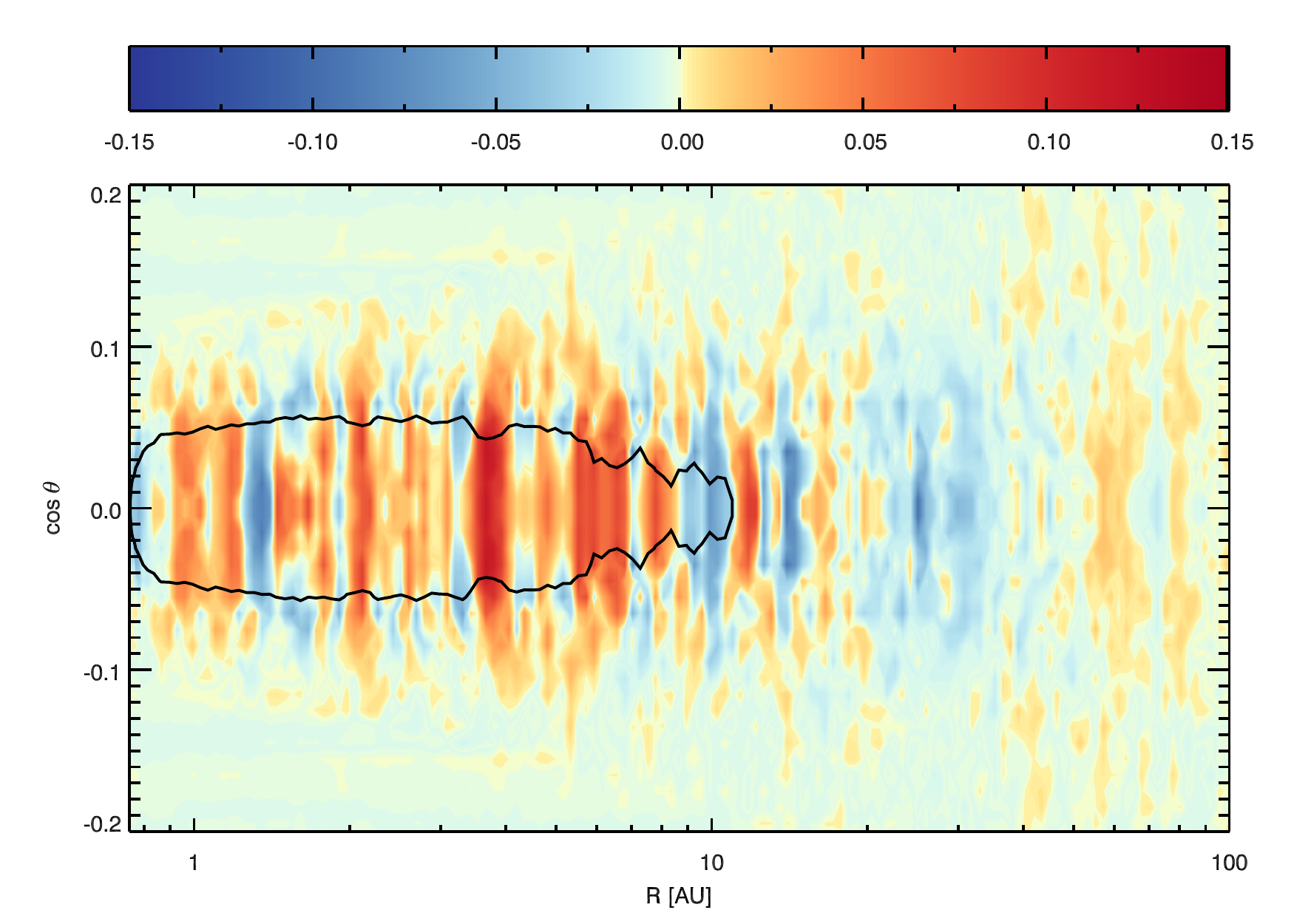}}
\caption{Same as the right panel of Fig.~\ref{fig:compare diffusion} but now comparing a model using $10^6$ photon packages with the reference model using $10^8$ photon packages.}
\label{fig:diffuse 1M}
\end{figure}

The errors when using only $10^5$ photon packages and the PDA are still considerable. This is because by using such low number of photons, the region where the diffusion approximation is employed is large. When we increase the number of photon packages to $10^6$ we can decrease the size of this region significantly (see Fig.~\ref{fig:diffuse 1M}). The maximum error in this case is $\sim12\%$, but only in a very small region of the disk. This model still runs in a few minutes on a normal desktop computer and because of the high gain in accuracy seems to be the best tradeoff between speed and accuracy for most cases.

\begin{figure*}[!t]
\resizebox{\hsize}{!}{\includegraphics{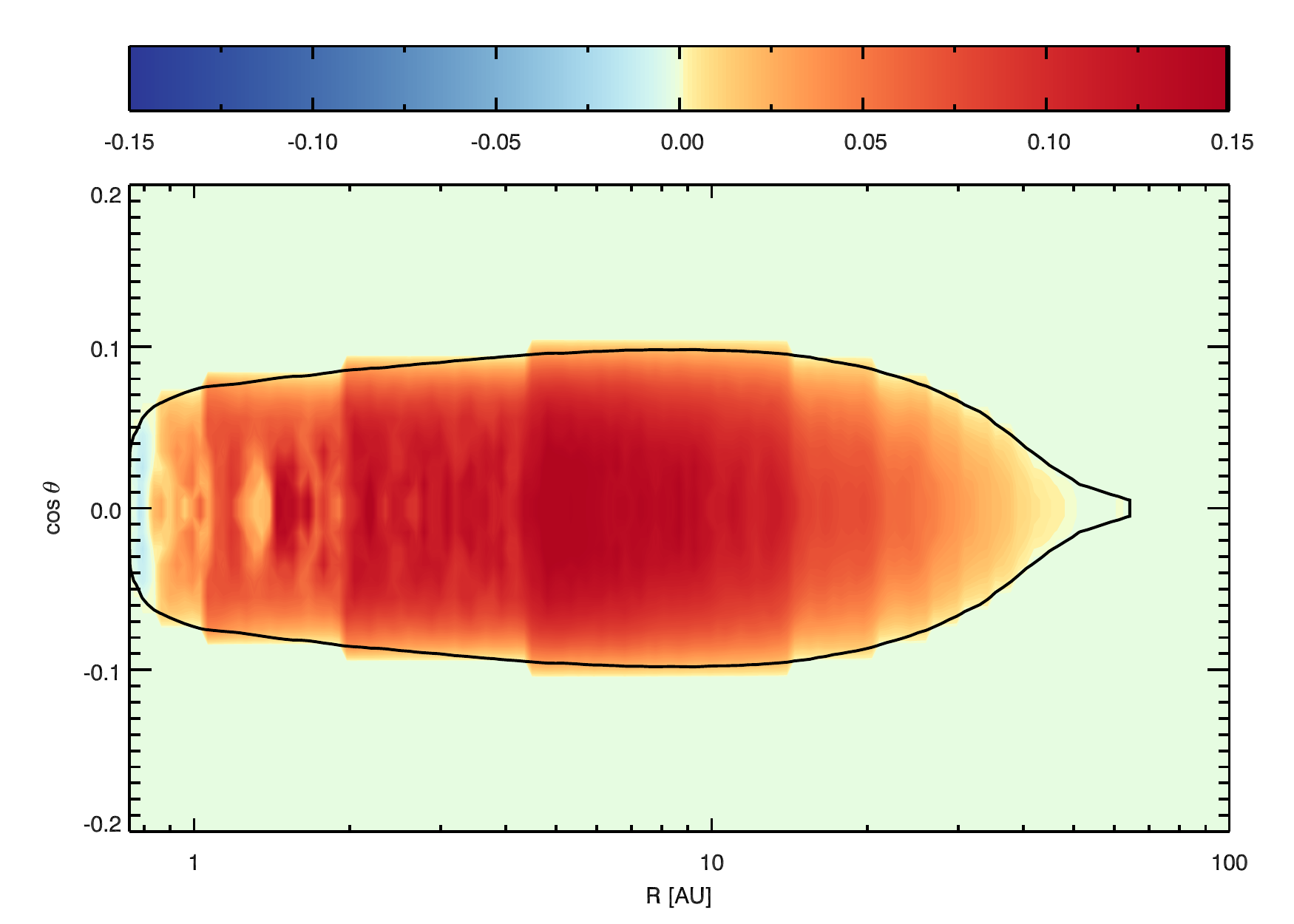}\includegraphics{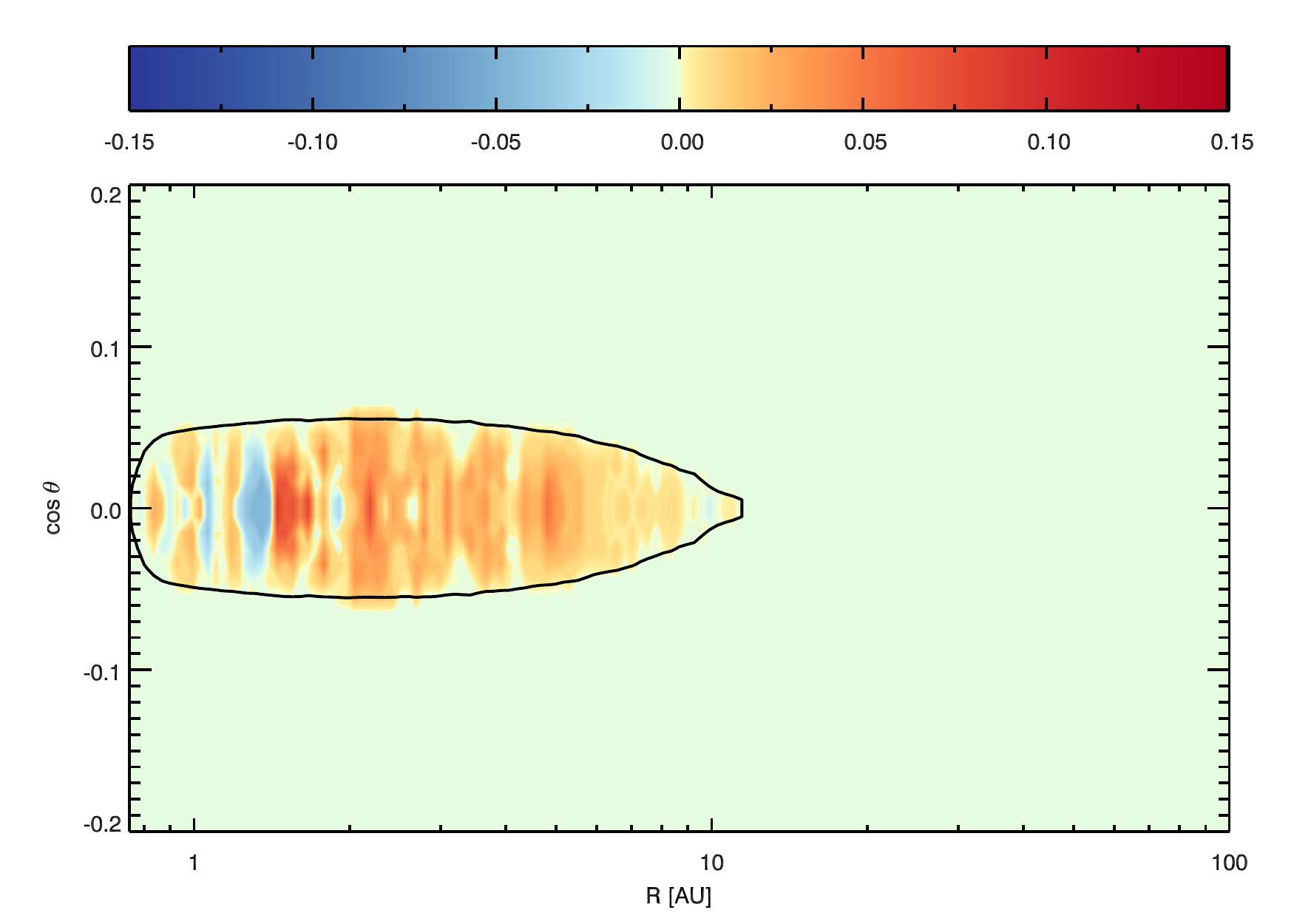}}
\caption{The relative error in the temperature computed by the model using $10^8$ photon packages with PDA for the region with 'low' photon counts as compared to the reference model using $10^8$ photon packages as a function of position in the disk. Everywhere in the disk where less than 30000 photon packages (left panel) or 3000 photon packages (right panel) entered we used the PDA. This region is enclosed by the black contour line. The dust mass in these models is $10^{-3}\,M_{\sun}$.}
\label{fig:compare diffusion2}
\end{figure*}

The diffusion approximation is only valid in regions with very high optical depths. The above errors are caused by the Monte Carlo noise and the intrinsic errors caused by the assumption entering the PDA. To study the size of the intrinsic errors we have used the partial diffusion approximation to compute the temperatures in the $10^8$ photon run in regions that have less than $30000$ or $3000$ photons. These regions are comparable to the regions where the PDA was used for the $10^5$ and the $10^6$ photon runs respectively. However, now the temperatures at the boundary of the diffusion region are determined to a very high accuracy. Therefore, we can now see the intrinsic error made by the PDA. For the region corresponding to the $10^5$ photon run (limit of 30000 photons) the largest relative error in the temperature compared to the reference model is still $\sim18\%$ 
(see Fig.~\ref{fig:compare diffusion2}). For the region corresponding to the $10^6$ photon run (limit 3000 photons) the largest relative error is $\sim8.5\%$.

As mentioned in section \ref{sec:Radiative diffusion} the PDA always tends to overestimate the temperature. In principle this causes violation of energy conservation. When the region where the PDA is used is sufficiently shielded from the observer, this is no problem. However, one can think of exotic density distributions where large, moderately optically thick regions are shielded from stellar radiation but not from the observer. In these cases, the errors in the temperature distribution can be high and will result in errors in the computed spectra and images when these are obtained from the obtained temperature distribution. This will be visible because in these cases the integrated total luminosity will be higher than the stellar luminosity. Since the PDA is only used after the Monte Carlo run, observables directly obtained from escaping photons will not be influenced. For ray-tracing the observables, like spectra, images and visibilities, the noise in the temperature structure usually provides no problems. Thus, when ray-tracing observables it is better to use the temperature as obtained before using the PDA. For determining the vertical scale height it is required that the temperature distribution is smooth and that the temperature is well defined throughout the entire disk. Therefore, temperatures obtained by the PDA are used when computing the vertical scaleheight of the disk (or other dynamical or geometrical properties of the disk that depend on temperature and require a smooth temperature profile).

\subsection{Results with modified random walk approximation}
\label{sec:random walk test}

\begin{figure*}[!t]
\resizebox{\hsize}{!}{\includegraphics{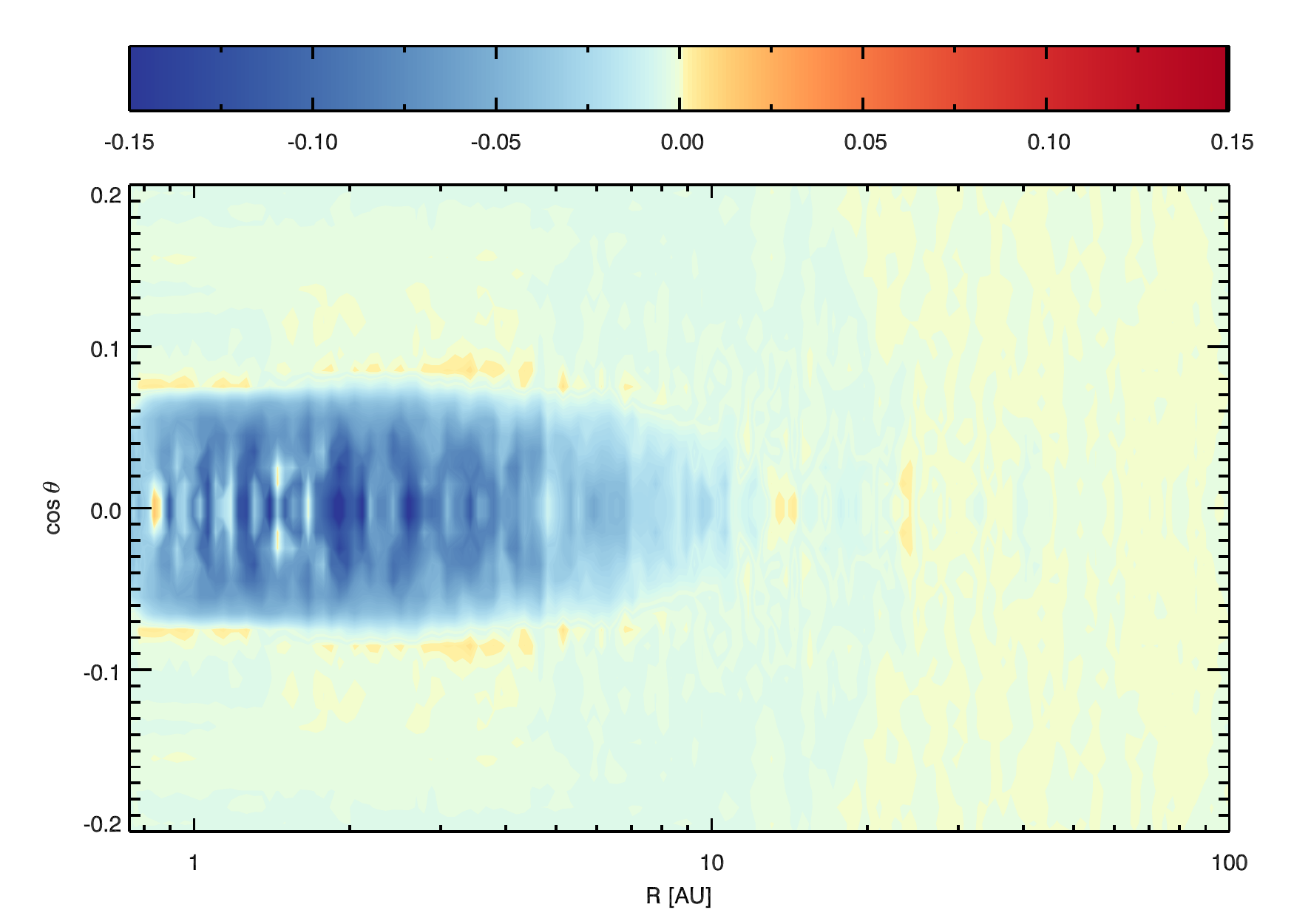}\includegraphics{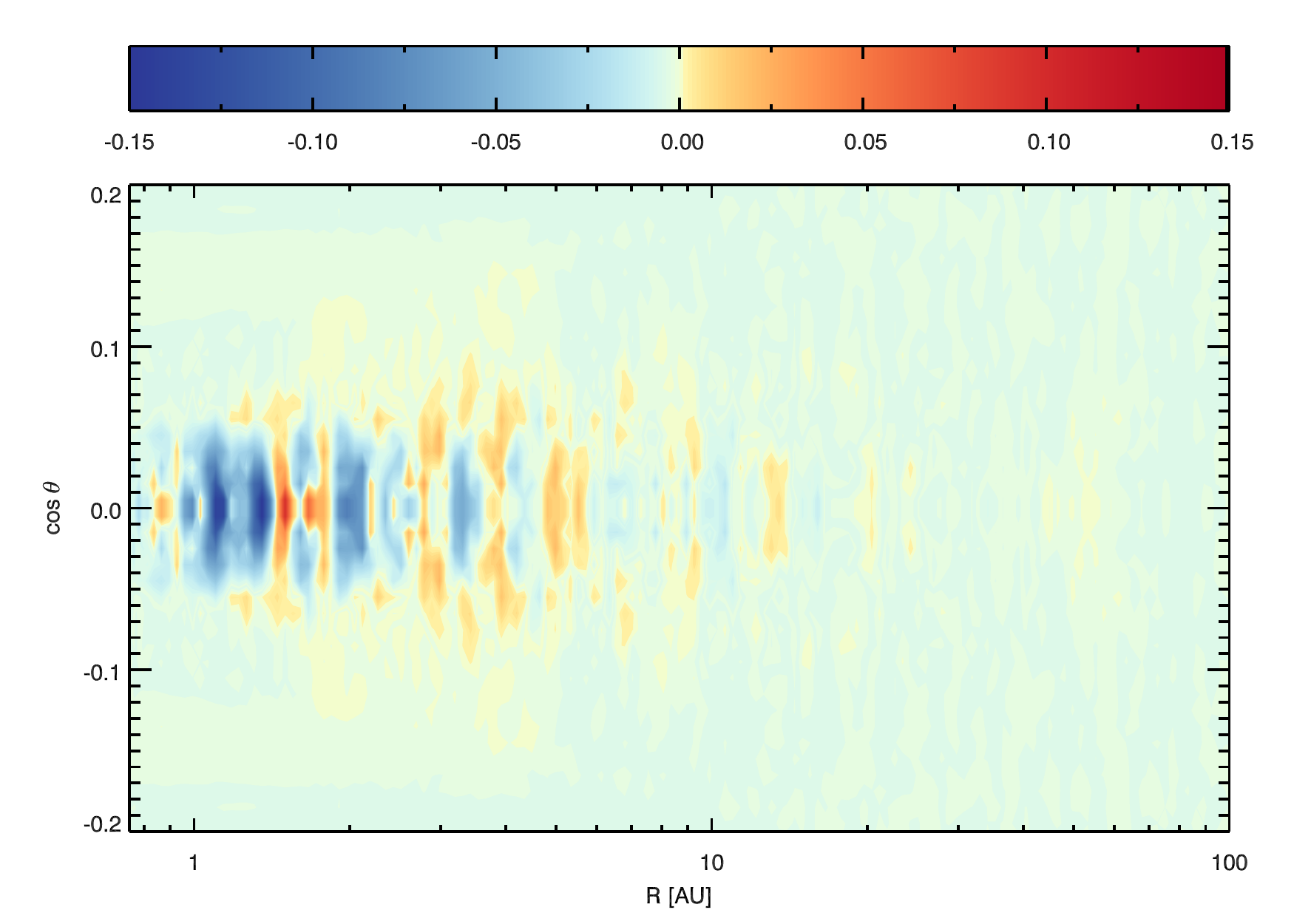}}
\caption{The relative error in the temperature using $10^8$ photon packages and the MRW compared to the reference model using $10^8$ photon packages as a function of position in the disk. The left panel shows the case where $\gamma=1$, the right panel shows the case for $\gamma=10$. The dust mass in these models is $10^{-3}\,M_{\sun}$.}
\label{fig:compare RW}
\end{figure*}

In this section we study the accuracy of the Modified Random Walk (MRW) procedure. We have run the radiative transfer code using $10^8$ photon packages with and without the MRW. We used $\gamma=1$ and $\gamma=10$. The difference in execution time is shown in Table~\ref{tab:timing}. In this table we present the average runtime per photon package for the various models with and without the modified random walk (MRW) method relative to the $10^{-6}\,$M$_{\sun}$ model without the MRW (which was 0.22\,ms in our case on a normal desktop computer). For this comparison we have also added disks with extreme mass ($M_\mathrm{dust}=10^{-2}\,M_{\sun}$). Although such disks are too massive to be stable in reality, it represents the results one would get for more extreme density distributions (for example a disk with the same mass but with a steeper power law index of the surface density). It is clear that the MRW speeds up the model for massive disks.

We compare the two cases with different values for $\gamma$ again with the reference model using $10^8$ photon packages. The results are shown in Fig.~\ref{fig:compare RW}. In the left panel we show the case for $\gamma=1$. The relative error in the temperature in this case varies between $-17\%$ and $+2\%$. The errors are concentrated in the innermost regions. The second case, with $\gamma=10$, is shown in the right panel of Fig.~\ref{fig:compare RW} and has comparable errors. In this case the errors vary between $-15\%$ and $+10\%$. More importantly, in this case the region where the errors occur is much smaller.

We conclude from the above that the temperatures are, on average, systematically underestimated by the MRW. Part of the reason for this is that for a small circumscribing sphere, the diffusion approximation used to derive the random walk procedure tends to underestimate the time it takes to leave this sphere. When the radius of the sphere is increased the approximation gets better. We can do this by increasing $\gamma$ which ensures that the MRW is only activated when the circumscribing sphere is large enough. For $\gamma=10$ we find that the errors are below $15\%$ everywhere in the disk, and that the region with the largest errors is very small and located near the midplane of the disk. A second reason for the errors on the temperature for low values of $\gamma$ is that the direction of propagation of the photon package after leaving the random walk sphere is not well defined. When the radius of the sphere is increased, the travel direction at the edge becomes less important and thus the temperature estimate becomes more accurate.

There is no effect of the errors on the temperature structure on the observables. This is because the regions where the MRW is activated are sufficiently shielded from the observer. However, note that in contrast to the PDA, the MRW is used \emph{during} the Monte Carlo procedure and thus, in principle, also influences the observables directly obtained from escaping Monte Carlo photon packages. These errors will be discussed below.

The MRW makes the dependence of the computation time on the mass of the disk much weaker. In addition to the cases presented in this paper, we have performed computations for extreme density distributions (e.g. with $\Sigma\propto r^{-2}$) where the decrease in computation time due to the random walk procedure is up to a few orders of magnitude.

\begin{table}[!t]
\caption{The relative average runtime per photon package for the various models.}
\begin{center}
\begin{tabular}{lccc}
\hline
$M_\mathrm{dust}$	& \multicolumn{3}{c}{Relative runtime per photon package}\\
$\left[M_{\sun}\right]$			& without MRW	& \multicolumn{2}{c}{with MRW} \\
					&			& $\gamma=1$		& $\gamma=10$ \\
\hline
$10^{-6}$			& 1		& 0.73			& 0.73 \\
$10^{-3}$			& 10.2	& 1.27			& 2.05 \\
$10^{-2}$			& 58.6	& 1.82			& 5.23 \\
\hline
\end{tabular}
\end{center}
\label{tab:timing}
\end{table}

\subsection{Observables: the effect on the SEDs}

To test the effects of the MRW on the predicted observables we have considered the spectral energy distribution (SED), visibility curves and images. We will discuss here only the results on the SEDs. For the case where we solve the hydrostatic equilibrium (section \ref{sec:vertical structure}) we will also show the resulting visibility curves. Note that the temperature computed using the PDA is not used in the ray-tracing procedure for computing the observables. Instead we use the temperature directly derived from the Monte Carlo run including the MRW procedure.

\begin{figure}[!t]
\resizebox{\hsize}{!}{\includegraphics{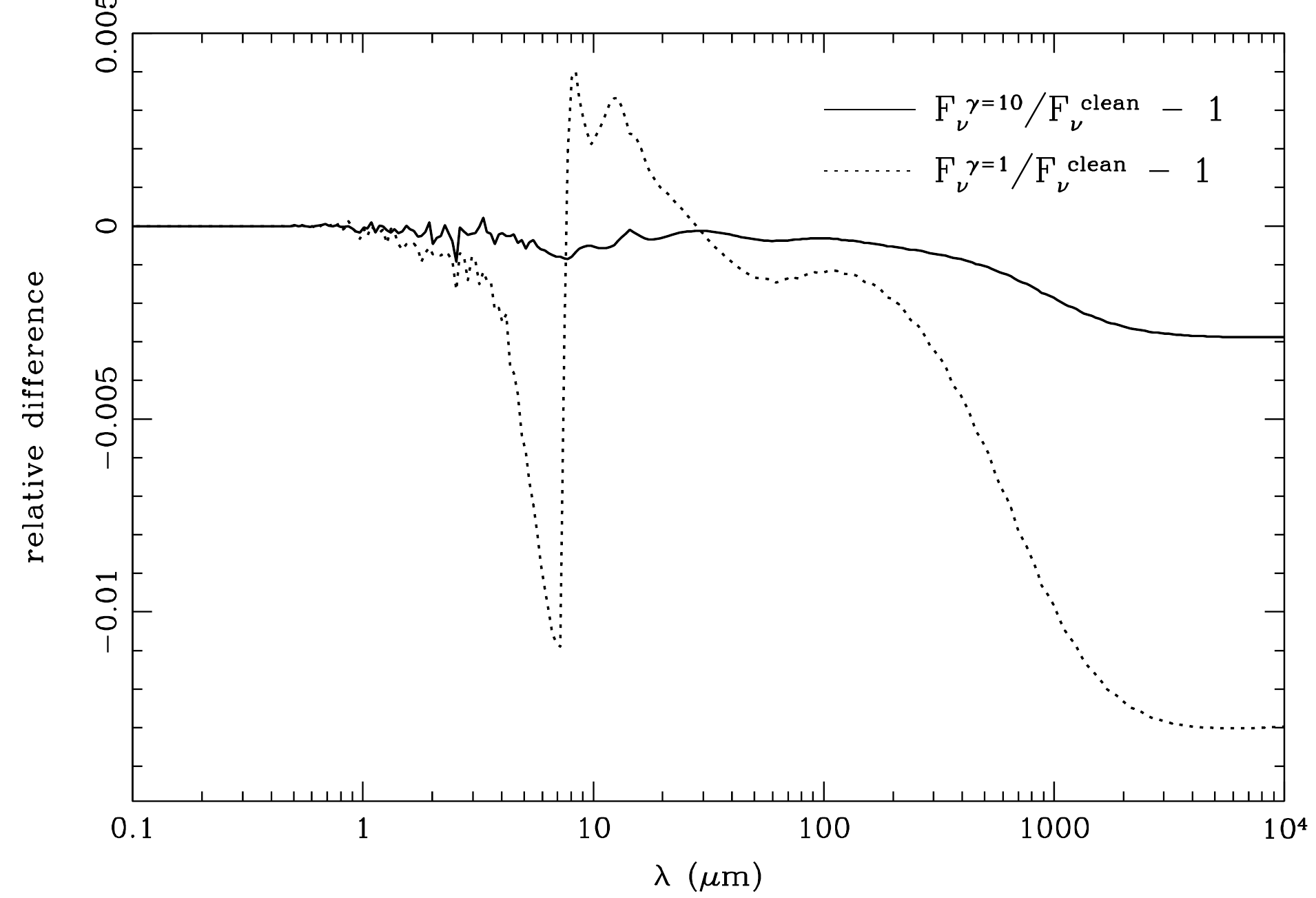}}
\caption{The relative differences between the SEDs computed with the random walk procedure using two different values of $\gamma$ as compared to the reference model without the random walk procedure. The results shown are for the high mass disk under an inclination of 45$^\circ$.}
\label{fig:error SED}
\end{figure}

For the SEDs we looked at the entire spectrum from 0.1$\,\mu$m up to 1$\,$cm. The resulting errors for the high mass disk seen under an inclination of 45$^\circ$ are plotted in Fig.~\ref{fig:error SED}. For the low mass disk the effects of using the random walk procedure is negligible, as it is only used in a very small region of the disk. For the high mass disk the effects of using the MRW with $\gamma=1$ can be as large as 1\% for a disk seen under moderate inclinations. For inclinations close to edge on (i.e. inclination higher than 80$^\circ$), the differences can be a few (up to 4) percent (not shown). For $\gamma=10$ the error made using the MRW is much smaller. For moderate inclinations it is always lower than 0.2\%. For nearly edge on disks the error can be up to 0.5\% (not shown). The increasing error at the long wavelength side of the SED is caused by the fact that at these wavelengths one starts to see through the disk, so the temperatures in the midplane are more important. The reason that the errors are much smaller in the case of $\gamma=10$ is that, although the errors on the temperature are of the same order, the region where this error is made is much smaller and deeper hidden in the disk. We conclude that the effect of the MRW procedure on the SED is very small.

\subsection{Solving the vertical density distribution}
\label{sec:vertical structure}

The vertical structure of the disk is mostly set by the midplane temperature, and therefore it is important to study the effects of the PDA and MRW approximations on the iterated vertical density distribution. We expect the effects of the errors of the temperature to translate into somewhat smaller errors in the vertical structure since the scale height of the disk is roughly proportional to $\sqrt{T}$. In this section we first analyse how the obtained structure of the disk is influenced, and after this evaluate what the consequences are for simulated observations. 
As a convergence criterion we use the difference between two iterations in terms of standard deviations, $\sigma$, integrated over the entire disk where we use the statistical error in the temperatures to compute a normalized standard deviation.

We consider here three cases:
\begin{description}
\item[\bf Case 1:] Using $10^5$ photon packages, applying the modified random walk ($\gamma=10$), and partial diffusion approximation (using a minimum number of photons of 30).
\item[\bf Case 2] Using $10^6$ photon packages, applying the modified random walk ($\gamma=10$), and partial diffusion approximation (using a minimum number of photons of 30).
\item[\bf Case 3] Using $2\cdot10^7$ photon packages, without applying the modified random walk. We do apply the partial diffusion approximation but with a minimum number of photons of only 1 (the partial diffusion approximation is only used to assure that no cells exist with a zero temperature, which would cause the vertical scale height to collapse).
\end{description}
Thus, here Case 3 is our reference model.

\begin{figure}[!t]
\resizebox{\hsize}{!}{\includegraphics{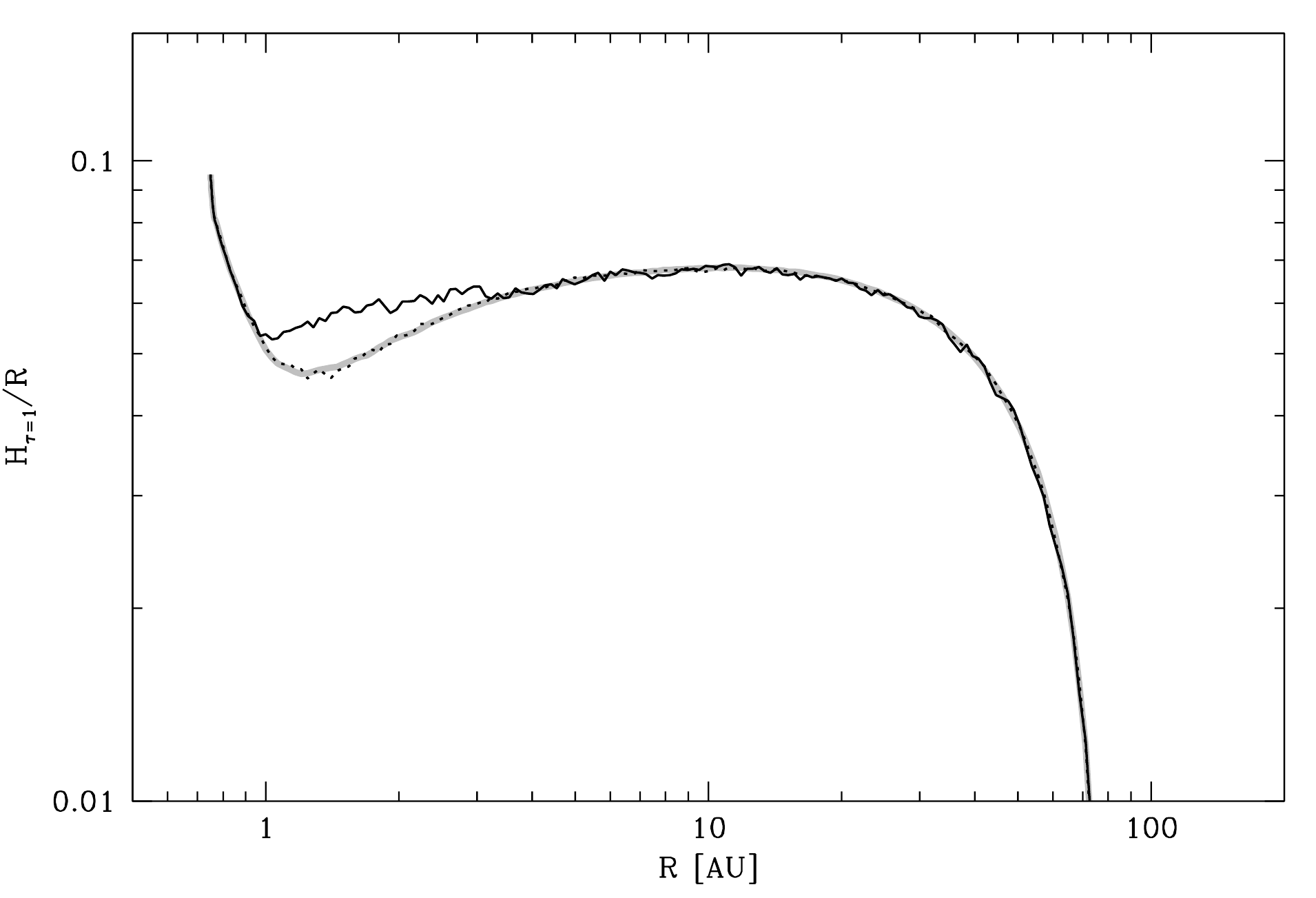}}
\caption{The vertical surface height of the low mass, i.e. $M_\mathrm{dust}=10^{-6}\,M_{\sun}$, disk. The ordinate is the height of the surface of the disk where an optical depth of unity is reached when looking from the top at a wavelength of $0.55\,\mu$m divided by the radius, $R$. All three cases described in the text are displayed. The solid black line represents Case 1, the dashed line Case 2, and the solid grey line Case 3.}
\label{fig:low mass}
\end{figure}

For the low mass disk ($M_\textrm{dust}=10^{-6}\,M_{\sun}$) the results of the vertical height of the disk is shown in Fig.~\ref{fig:low mass}.  In this figure we show the height of the surface of the disk where an optical depth of unity is reached in the direction perpendicular to the midplane at a wavelength of $0.55\,\mu$m. In the first case (using $10^5$ photon packages) the diffusion region used for the PDA is quite large, causing an overestimate of the midplane temperature and therefore an overestimate of the height of the disk above the diffusion region. This region lies just behind the inner rim. By using $10^6$ photon packages (Case 2) the diffusion region is sufficiently small to overcome this problem, and no significant differences are observed between Case 2 and 3. In all cases convergence was reached within 9 iterations.

\begin{figure}[!t]
\resizebox{\hsize}{!}{\includegraphics{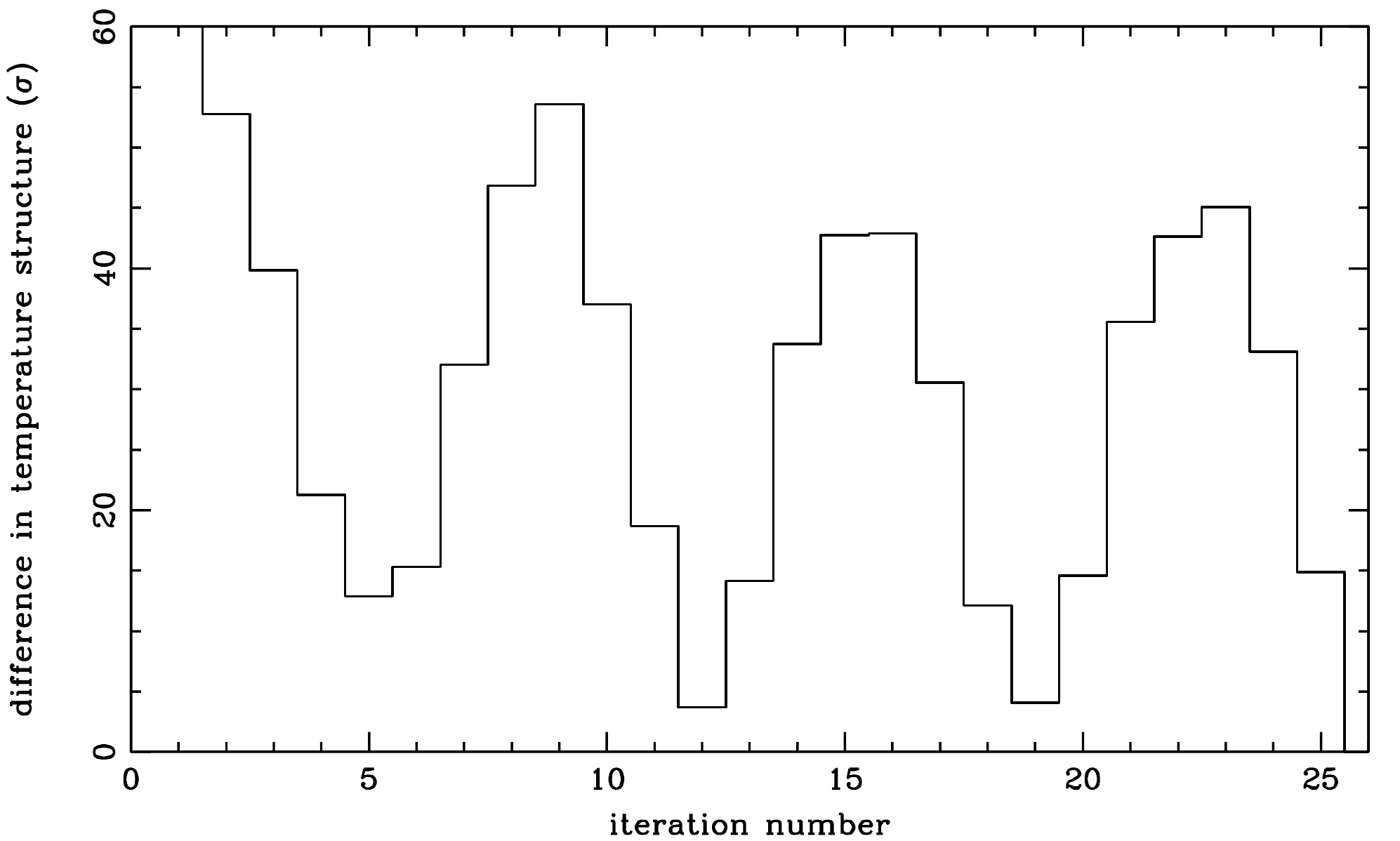}}
\caption{The differences in terms of $\sigma$ (see text) between the temperature structures computed in the 26th iteration and all previous iterations. Clearly, the structure as in the 26th iteration is close to that already computed in the 19th and 12th iteration. This oscillatory behavior is caused by 'waves' propagating along the surface of the disk.}
\label{fig:high mass fluct}
\end{figure}

\begin{figure}[!t]
\resizebox{\hsize}{!}{\includegraphics{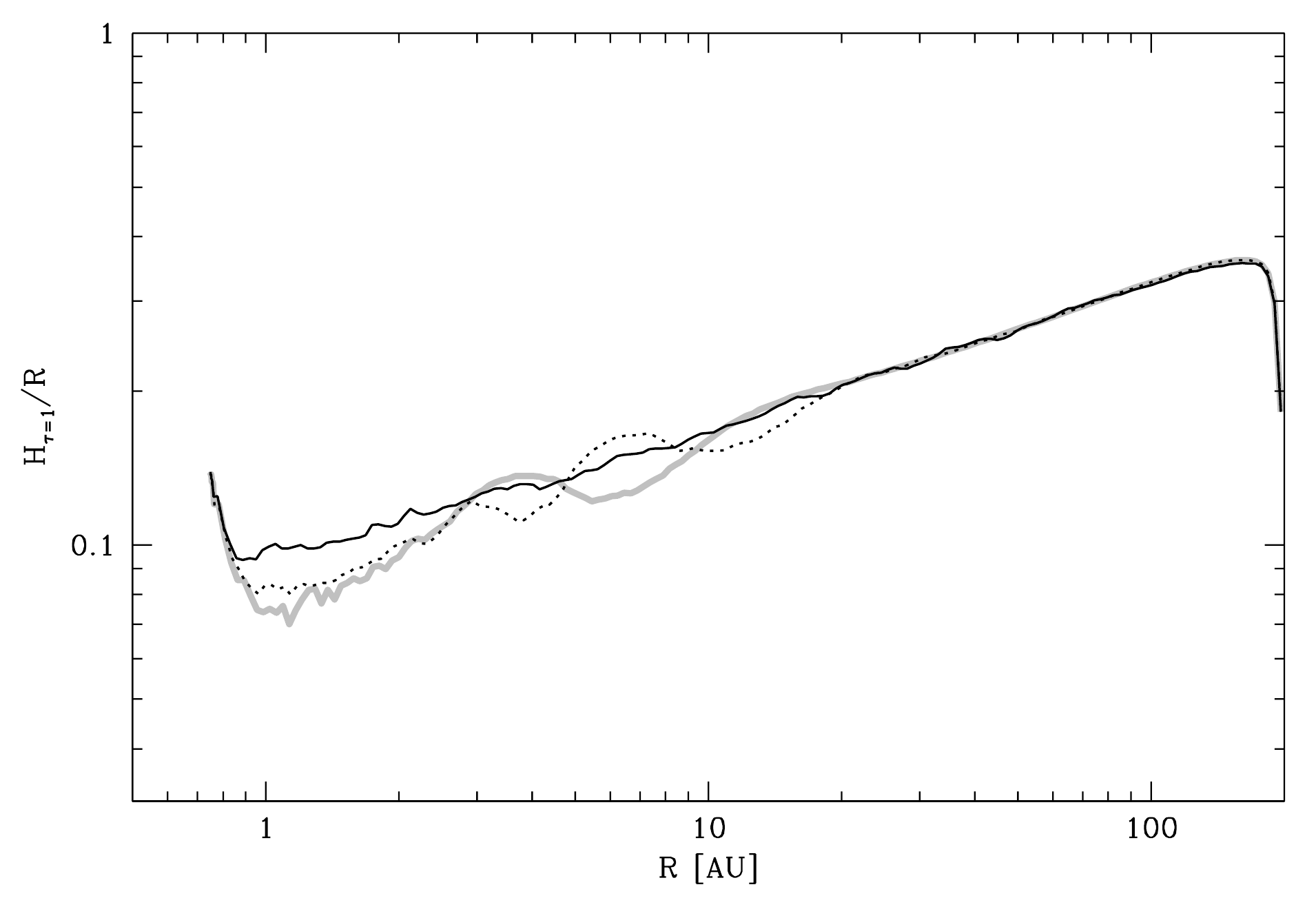}}
\caption{Same as Fig.~\ref{fig:low mass} but for a dust mass of $10^{-3}\,M_{\sun}$. The solid black line represents Case 1, the dashed line Case 2, and the solid grey line Case 3. The curves for Case 2 and 3 represent oscillatory solutions (see main text).}
\label{fig:high mass}
\end{figure}

For the higher mass disk it is much harder to reach convergence. For this disk mass, for both Case 2 and 3 periodic fluctuations are observed with a period of 7 to 8 iterations (see Fig.~\ref{fig:high mass fluct}). These fluctuations are caused by instabilities in the disk structure as described by \citet{1999ApJ...511..896D, 2000A&A...361L..17D, 2008ApJ...672.1183W} and cause the convergence criterion to fail. As can be seen from Fig.~\ref{fig:high mass fluct} the solution is periodic, implying that the procedure is converged but to an oscillatory solution. The question currently remains if in reality these fluctuations are suppressed by viscous heating, turbulence or hydrodynamical effects causing the disk to respond slower than the heating and cooling timescale. For Case 1 ($10^5$ photon packages) these fluctuations are not observed and convergence is reached in 7 iterations (see Fig.~\ref{fig:high mass}). However, the resulting vertical structure is too high in the region just behind the inner rim. This is, again, caused by the overestimate of the temperature at the midplane due to the PDA. We conclude that for both cases using $10^6$ photon packages is sufficient to obtain a reliable vertical density profile. In this case the diffusion region, where the PDA is employed, is deep enough in the disk to ensure the diffusion approximation to be valid and not cause large errors.

\begin{figure*}[!t]
\centerline{\resizebox{13cm}{!}{\includegraphics{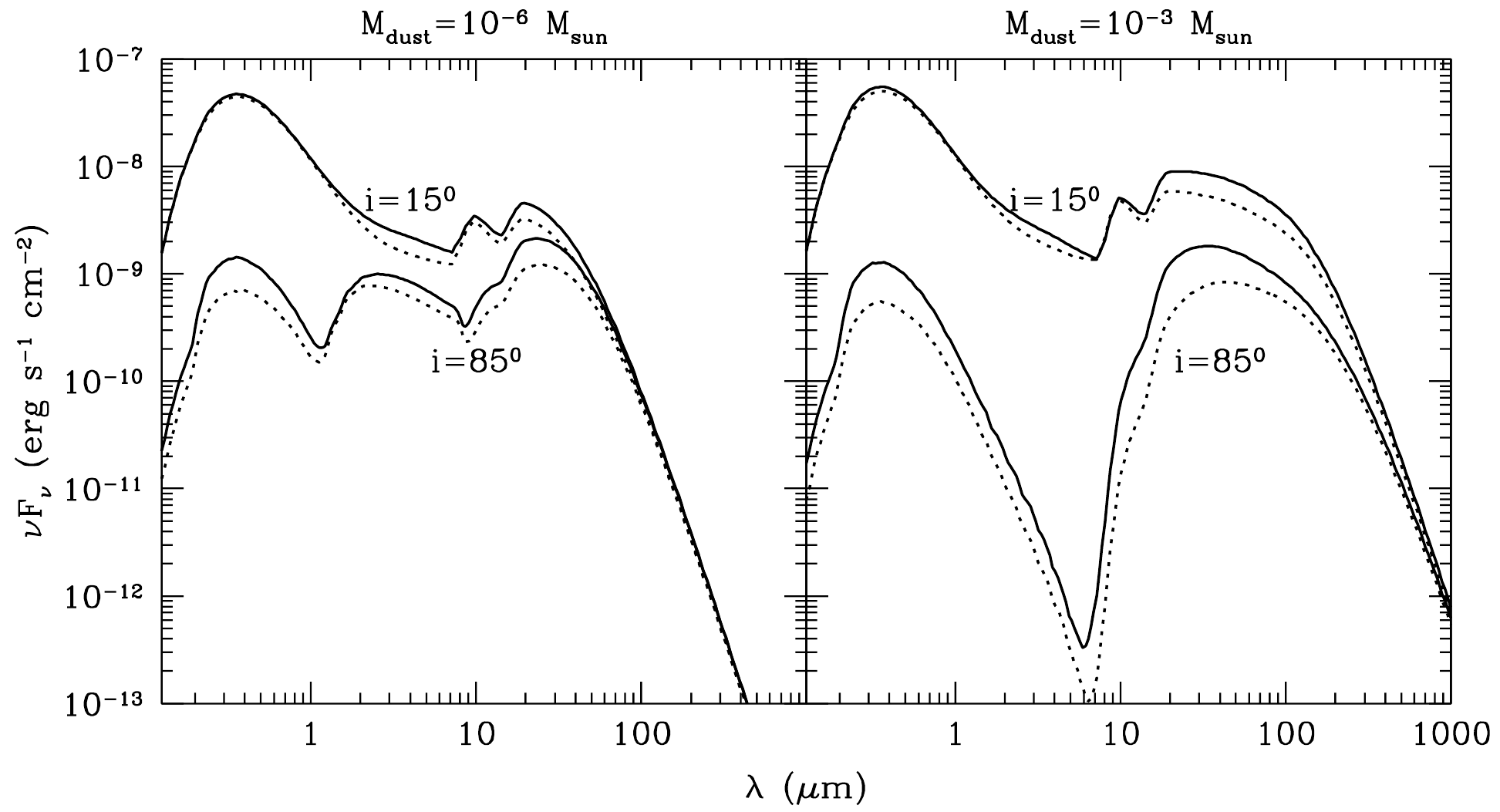}}}
\caption{The resulting SEDs as computed for the two disk masses after iteration on the vertical structure (solid lines). The corresponding SEDs for the parameterized disk structures are also shown (dotted lines). The distance to the object is 150\,pc.}
\label{fig:spectra iter}
\end{figure*}

\begin{figure}[!t]
\resizebox{\hsize}{!}{\includegraphics{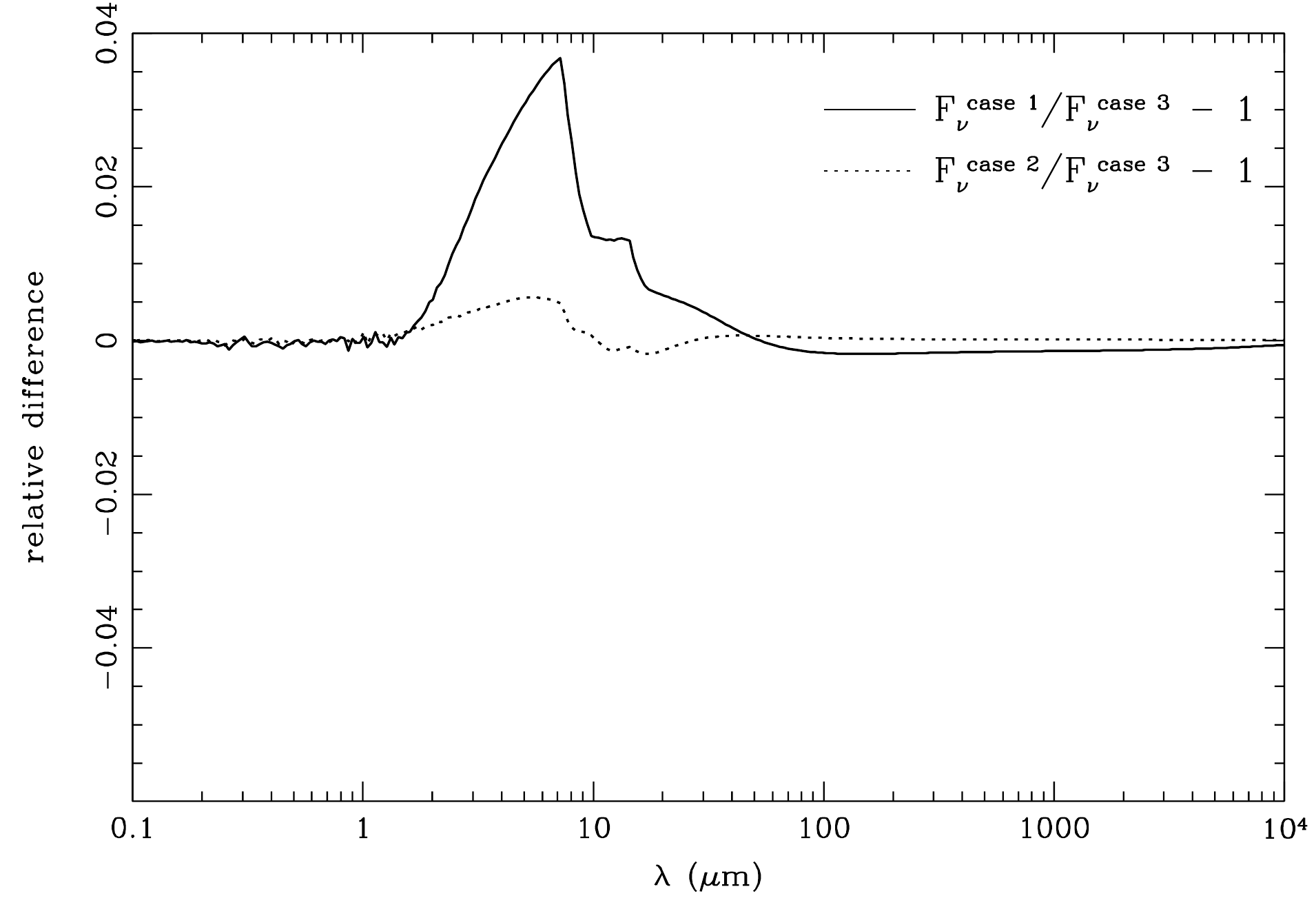}}\\
\resizebox{\hsize}{!}{\includegraphics{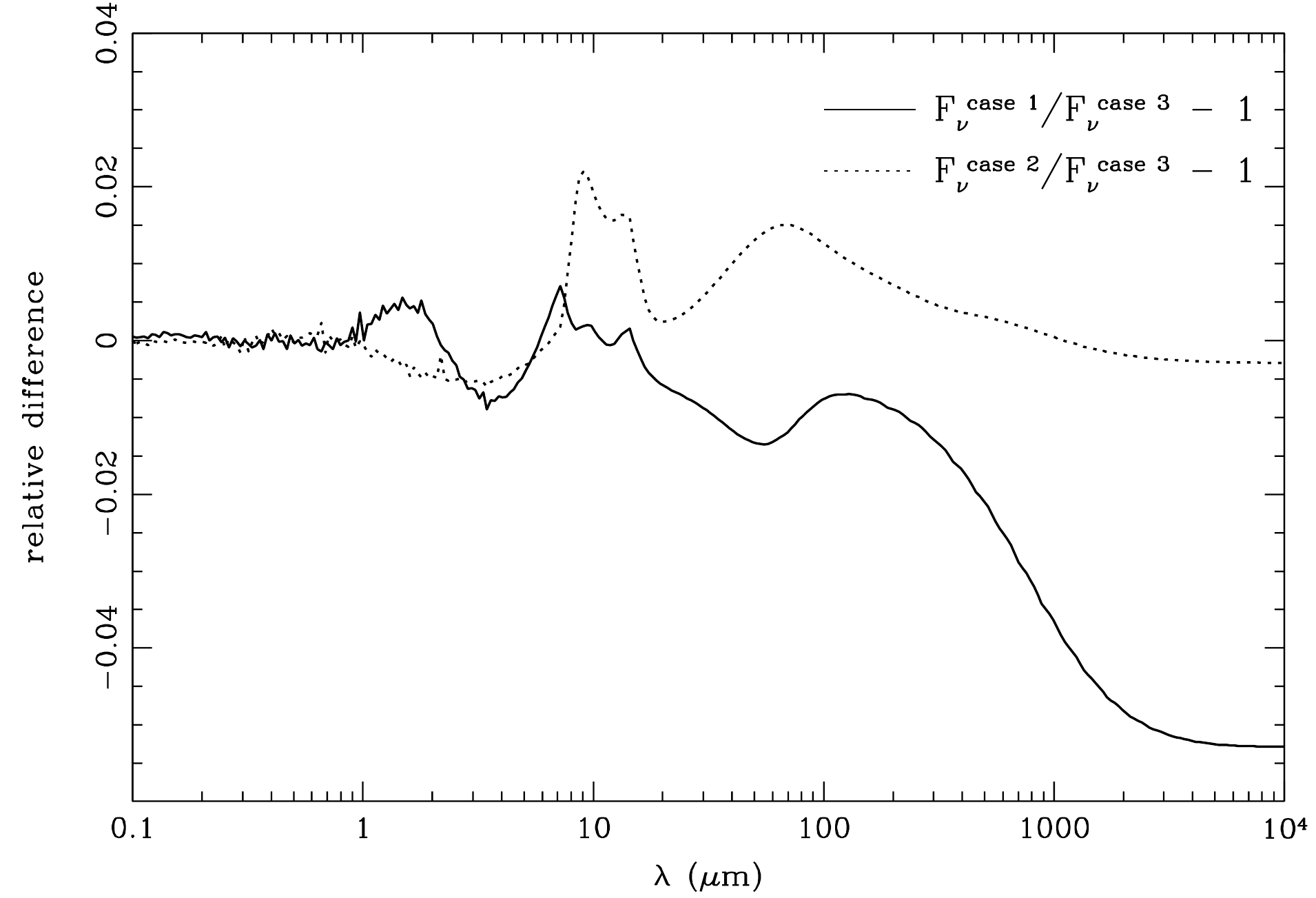}}
\caption{The relative differences between the SEDs computed for the three different cases defined in the text. The upper panel shows the results for the dust mass of $10^{-6}\,M_{\sun}$, the lower for a dust mass of $10^{-3}\,M_{\sun}$.}
\label{fig:error SED iter}
\end{figure}

We have computed the spectra and visibility curves of the resulting density and temperature structures. The spectra of the iterated disks (case 1) together with those of the corresponding parametrized disks are shown for disks viewed at an inclination of 45$^\circ$ in Fig.~\ref{fig:spectra iter}. In order to judge the effects of the overestimate of the disk height just behind the inner rim we show the differences in the SEDs between the different cases we considered in Fig.~\ref{fig:error SED iter} for a disk viewed at an inclination of 15$^\circ$ and 85$^\circ$. For the low mass case it is clear that the largest error is made in the 1-10\,$\mu$m region. The emission in this spectral region predominantly comes from those radii in the disk where the height of the disk is increased because of the use of the PDA (see Fig.~\ref{fig:low mass}), causing these regions to emit more efficiently when the PDA is employed. The errors at long wavelengths, as seen in the high mass case, are caused by the use of the MRW, as discussed above for the non-iterated case. In all cases we studied the differences are smaller than 5\%, independent of disk inclination. Therefore, we conclude that for modeling SED observations, it suffices to use the very fast, i.e. $10^5$ photon package, method with the PDA. Also, the effects of the oscillations in the vertical structure of the massive disk have only a very small impact (less than 5\%) on the SEDs.

\begin{figure}[!t]
\centerline{\resizebox{\hsize}{!}{\includegraphics{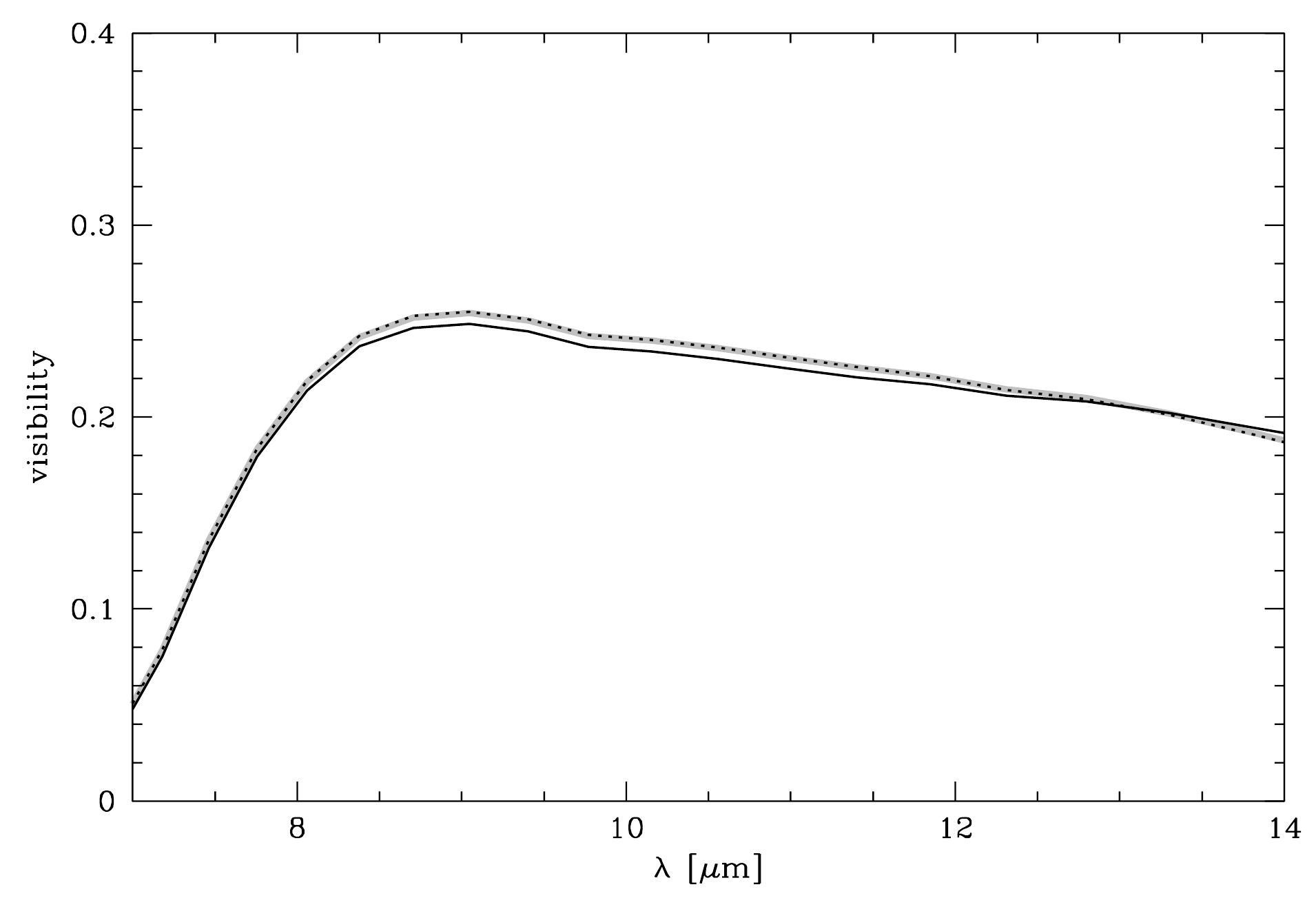}}}
\caption{The visibility curves for the low mass, i.e. $M_\mathrm{dust}=10^{-6}\,M_{\sun}$, disk and a baseline of 100 meter. The distance to the object is 150\,pc. All three cases described in the text are shown. The solid black line represents Case 1, the dashed line Case 2, and the solid grey line Case 3.}
\label{fig:visibilities}
\end{figure}

\begin{figure}[!t]
\centerline{\resizebox{\hsize}{!}{\includegraphics{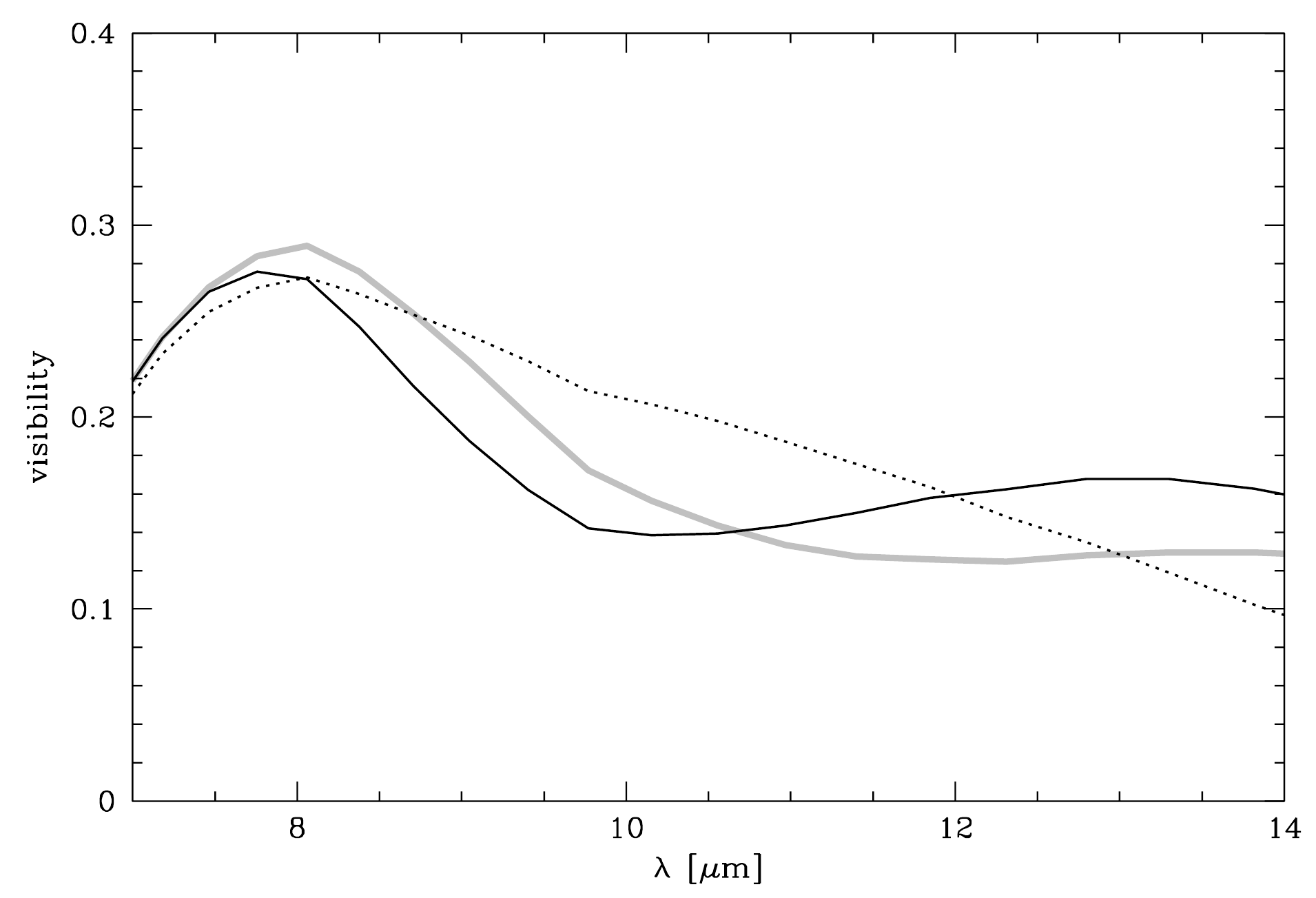}}}
\caption{The visibility curves for the high mass, i.e. $M_\mathrm{dust}=10^{-3}\,M_{\sun}$, disk and a baseline of 100 meter. The distance to the object is 150\,pc. The visibilities computed after three different numbers of iterations for Case 2 are shown. The solid black line is after 19 iterations, the dashed line after 23 iterations, and the solid grey line after 26 iterations.}
\label{fig:visibilities osc}
\end{figure}

For the visibilities things can be different. In Fig.~\ref{fig:visibilities} we show the visibility curves in the N-band, as would be observed by e.g. the Mid-Infrared Interferometric Instrument (MIDI) at the Very Large Telescope Interferometer (VLTI), for the low mass disk at a baseline of 100 meters (the disk is set at a distance of 150\,pc). It is already clear that the overestimate in the vertical height for Case 1 results in a slightly different visibility profile. For the high mass disk we have no convergence of the vertical structure. For Case 1 we get very different results as compared to either Case 2 or 3 which give visibility curves that are very close together (not shown). In Fig.~\ref{fig:visibilities osc} we show the visibility curves for Case 2 after different numbers of iterations. It is clear that the oscillatory behavior seen in the surface height of the disk can be observed in the visibility curves.

The oscillations found in the vertical structure of the disk do not appear to be caused by numerical problems. The waves originate in the region that just emerges from the shadow of the inner rim, and is thus again directly illuminated by the central star. In our case this is around 8-9\,AU. From there onwards they travel in until they fade out inside the shadow of the inner rim. The fact that it doesn't seem to be a numerical artifact means that if there are no processes to stabilize the vertical structure \citep[like those discussed by][]{1999ApJ...511..896D, 2000A&A...361L..17D}, the oscillations in the visibilities mentioned above could in principle be observed. The period with which these oscillations occur can be estimated from the time it takes for a sound wave to propagate from the midplane to the surface of the disk. Around 8\,AU, where the waves originate, this timescale is $\sim2$ years. The period of the oscillations is around 7 iterations and thus in time is estimated to be roughly 7 times the vertical timescale, i.e. $14\,$years. We should note that the timescale for a wave to travel from the midplane to the surface of the disk is a function of the distance to the central star. For example, at 3\,AU the timescale is only $\sim5$ months. It is hard to estimate from simple considerations if this will have an increasing or a decreasing effect on the strength of the waves. Computations using radiation hydrodynamics have to be performed to study these effects in more detail.

\section{Recommendations for use and implementation}
\label{sec:recommendations}

In this section we summarize the results discussed above and make recommendations for use and implementation of the PDA and the MRW procedures for various cases.

\subsection{Passively heated disks}

For radiative transfer through a fixed density structure the recommended setup for computations depends on the observables one wants to obtain. For computing SEDs or optical/infrared images it suffices to use only the MRW procedure, the PDA is then not required. Since the errors made by the MRW procedure are localized strongly in the high density regions, one can use a low value of $\gamma$ (e.g. $\sim1-5$). Errors are then smaller than $1.5$\%. When also millimeter images or visibilities are required a higher value of $\gamma$ is recommended (e.g. $\sim10$).

When accurate temperatures are required, e.g. for computations of chemical reactions or of the vertical structure, the PDA can be switched on. The number of photon packages used and the limit when the PDA is used can be tuned to the desired accuracy and computational speed. One needs to keep in mind that the diffusion approximation itself is not without error, so the use of a sufficient number of photon packages is advised. In order to keep the statistical error in the temperature structure below $\sim5$\% a lower limit of the number of photon packages per cell has to be set to approximately 30.

There are roughly two types of situations which require different considerations for the limits of the diffusion approximation and the number of photons needed which we will refer to as \emph{local accuracy} and \emph{global accuracy}.

\subsubsection{Local accuracy:} The first, and most difficult case is when the temperature is needed at each location in the disk with high accuracy, e.g. $5$\%. In this case, the PDA has to be restricted to a very small region deep inside the disk to decrease the intrinsic error of the diffusion approximation. In order to do so, a sufficient number of photon packages (e.g. $\sim10^7$) have to be used, and the limiting number of packages has to be set to $\sim30$ to avoid statistical errors.

\subsubsection{Global accuracy:} The second situation is when it is most important that the temperature structure is not too noisy, i.e. the fluctuations from cell to cell need to be small. This can be the case when one wants to look at more global disk characteristics in which the temperature plays a role like, for example, when solving for the vertical structure of the disk. In this case, a larger error in the temperature can be allowed in favor of a more stable and smooth temperature structure. Consequently, fewer photon packages can be used (e.g. $\sim10^6$) and the limit for using the PDA can be set to $\sim30$. Although the errors on the temperature structure can in this case be about 10\%, the noise level is much lower due to the smoothing caused by the PDA. Also, in this case the $\gamma$ for the MRW procedure can be set to a relatively low value (e.g. $\gamma=1$) since the temperatures in the region where the MRW is actually employed will be largely overwritten by the PDA.

\subsection{Viscously heated disks}

Although in this paper we have not considered the case of viscous heating, the MRW procedure does allow for complete transport of the energy released by viscous heating through the disk. When the MRW is not employed it is not feasible to transport a significant number of photon packages from the midplane, where the energy is released, to the surface layers. The computations we performed on the high mass disk showed that without the MRW procedure the computation time for a photon package to reach the surface layers was roughly a minute on a normal desktop computer. By using the MRW procedure with $\gamma=1$ this was reduced, in our case, to less than a second. This still does not allow for a very large number of photon packages, but at least several thousands can be used. The energy released by viscous heating ensures that the regions shielded from stellar radiation also receive sufficient photon packages to determine an accurate temperature. Thus in this case the PDA can be turned off. More details on this subject are beyond the scope of this paper and will be left for a future study.

\section{Conclusions}
\label{sec:conclusions}

We have presented two efficient implementations of the diffusion approximation in the high density regions of massive protoplanetary disks: the partial diffusion approximation and the modified random walk procedure.

First, the partial diffusion approximation (PDA) can be used to increase the accuracy of the temperature structure in the highly obscured regions. The PDA is mainly useful when the vertical scaleheight of the disk has to be computed self-consistently. In order to converge the vertical scaleheight computations it is important to have a temperature structure that is stable throughout the disk. The partial diffusion approximation is therefore ideal to reduce the photon noise, inherent to Monte-Carlo radiative transfer computations, in the midplane regions.

Second, the modified random walk (MRW) procedure increases the computational speed significantly, while errors are small. When the modified random walk procedure is not employed the computation time required is a strong function of the mass of the disk. When it is employed, this dependence is much weaker allowing for computations of disks with much more extreme optical depths, like for example in disks with steeper density gradients.

The combination of the above two methods allows one to perform computations for massive protoplanetary disks with high accuracy and within reasonable time. The modified random walk procedure ensures that enough photons can be emitted in order to get reasonable photon counts in most of the disk. The partial diffusion approximation can then be employed to correct the temperatures in those regions where errors are still large due to poor statistics.

We have performed computations to iteratively compute the vertical scaleheight of the disk. For this we have studied carefully the effects of both approximations mentioned above. The effects of using the MRW (with $\gamma=10$) on the vertical structure are very small, and have no impact on the observables. When very few photon packages are used, the region where one has to apply the PDA is large. This causes overestimates of the temperatures in a relatively large region of the disk, resulting in a local overestimate of the vertical height. Although this has no effect on the SED, the visibilities obtained with interferometric measurements are affected. We conclude that for computations of spatially unresolved observations, only a small number of photon packages is enough. When spatially resolved observations are to be simulated, one needs to resort to higher numbers of photon packages in order to shrink the region where the PDA has to be applied.

We also find that the vertical structure of massive disks converges to an oscillatory solution. 'Waves' traveling the surface of the disk cause the density structure to fluctuate. These waves originate at the location where the disk emerges from the shadow of the inner rim and from there travel inwards. When the damping mechanisms for these waves are not efficient, this can be important when considering spatially resolved computations or observations of flaring protoplanetary disks.

\begin{acknowledgements}
M.  Min acknowledges financial support from the Netherlands Organisation for Scientific Research (NWO) through a Veni grant. We would like to thank an anonymous referee for positive and constructive feedback on the paper.
\end{acknowledgements}

\appendix
\section{Benchmark test of MCMax}
\label{app:accuracy}

The new code MCMax has been tested for accuracy. We did this by comparing the computed spectra and temperature structures to those obtained by the independently developed radiative transfer code RADMC \citep[first used in][]{2004A&A...417..159D}. Both codes have been checked to reproduce the benchmark results of \citet{2004A&A...417..793P}. In addition to this both codes are involved in benchmark computations for higher mass disks by Pinte et al. (\emph{in prep}) and no discrepancies with other codes have been found so far.

We tested the codes MCMax and RADMC first for the benchmark setup of \citet{2004A&A...417..793P}. All results where reproduced without any problems. The differences with the benchmark SEDs are comparable with those presented by \citet{2004A&A...417..793P} between the different codes used in the benchmark computations. For the temperature structures we compared the results with those computed using the RADICAL code and found that the differences are in all cases less than $2\%$.

As an additional test case we performed computations for the $10^{-3}$ and $10^{-6}$ solar mass dust disks given the density setup as described in section~\ref{sec:disk setup} using $10^8$ photon packages for both codes. By using this number of photon packages we ensure that the photon count is very high everywhere in the disk. For this run we did not use the random walk or the diffusion approximation to ensure that the results are not influenced by the assumptions made in these approximations.

\begin{figure*}[!t]
\resizebox{\hsize}{!}{\includegraphics{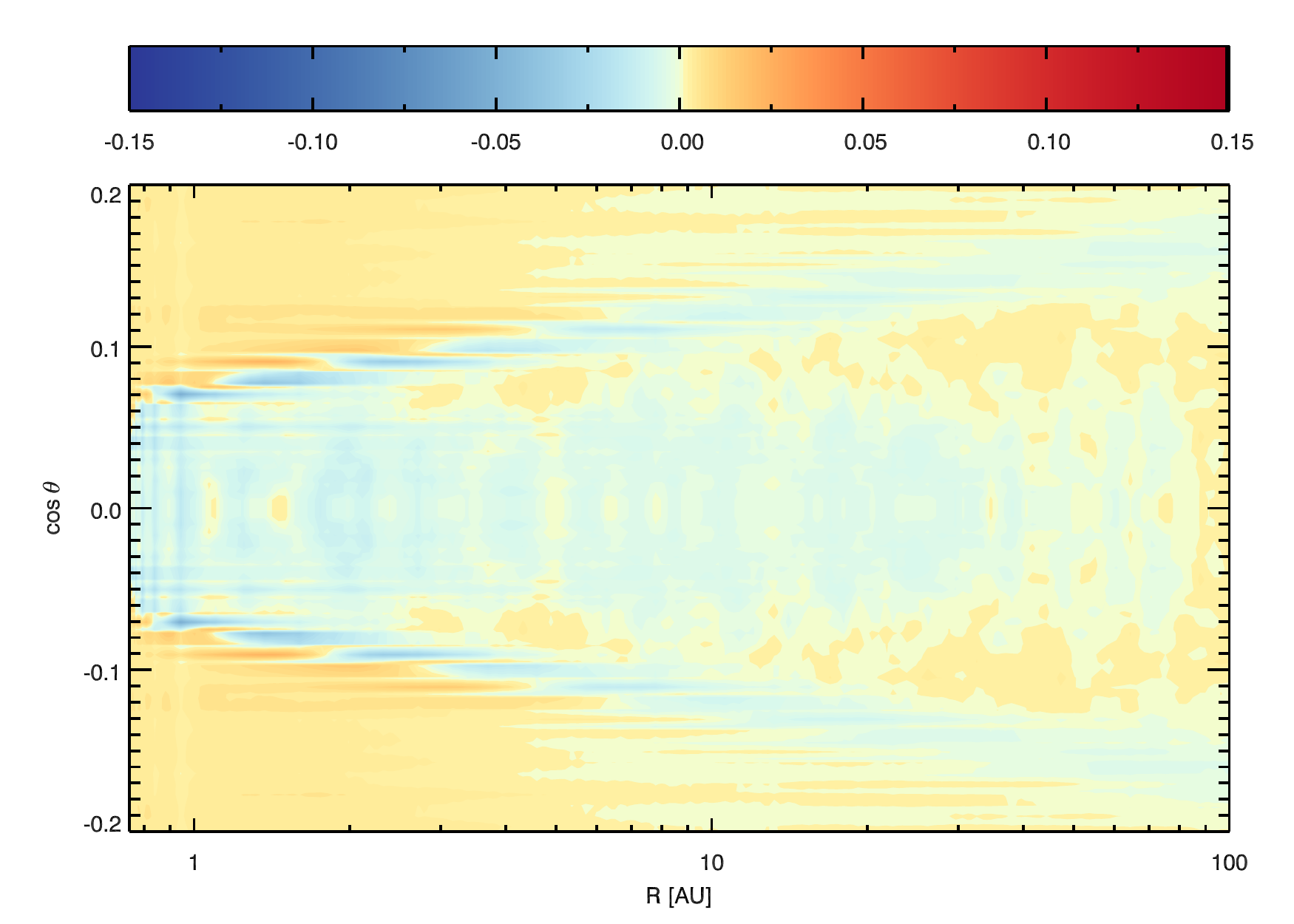}\includegraphics{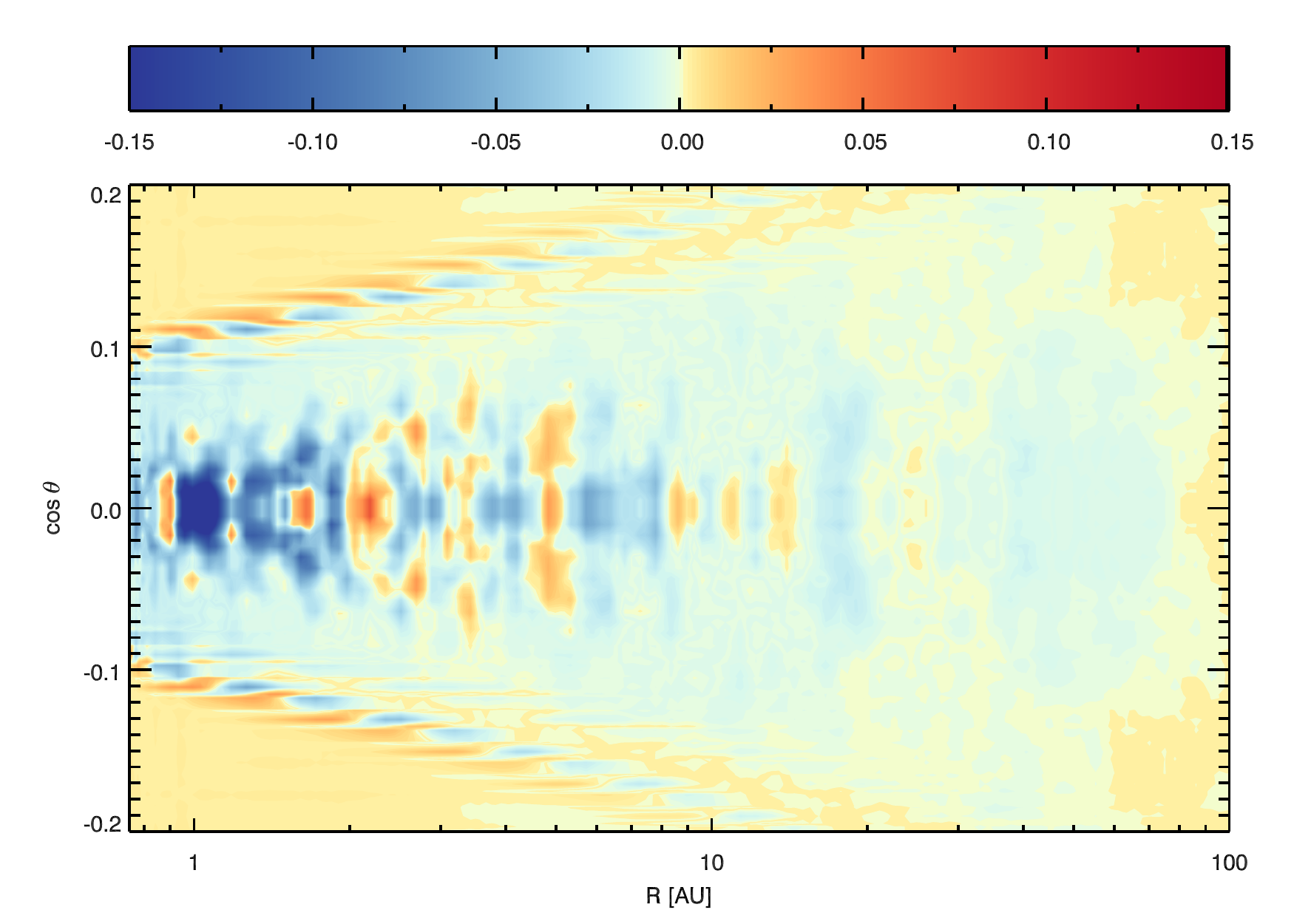}}
\caption{The relative error in the temperature computed by MCMax and RADMC. The left panel shows the model with a dust mass of $10^{-6}\,M_{\sun}$, while the right panel shows the model with $10^{-3}\,M_{\sun}$ of dust.}
\label{fig:compare codes}
\end{figure*}

\begin{figure*}[!t]
\centerline{\resizebox{13cm}{!}{\includegraphics{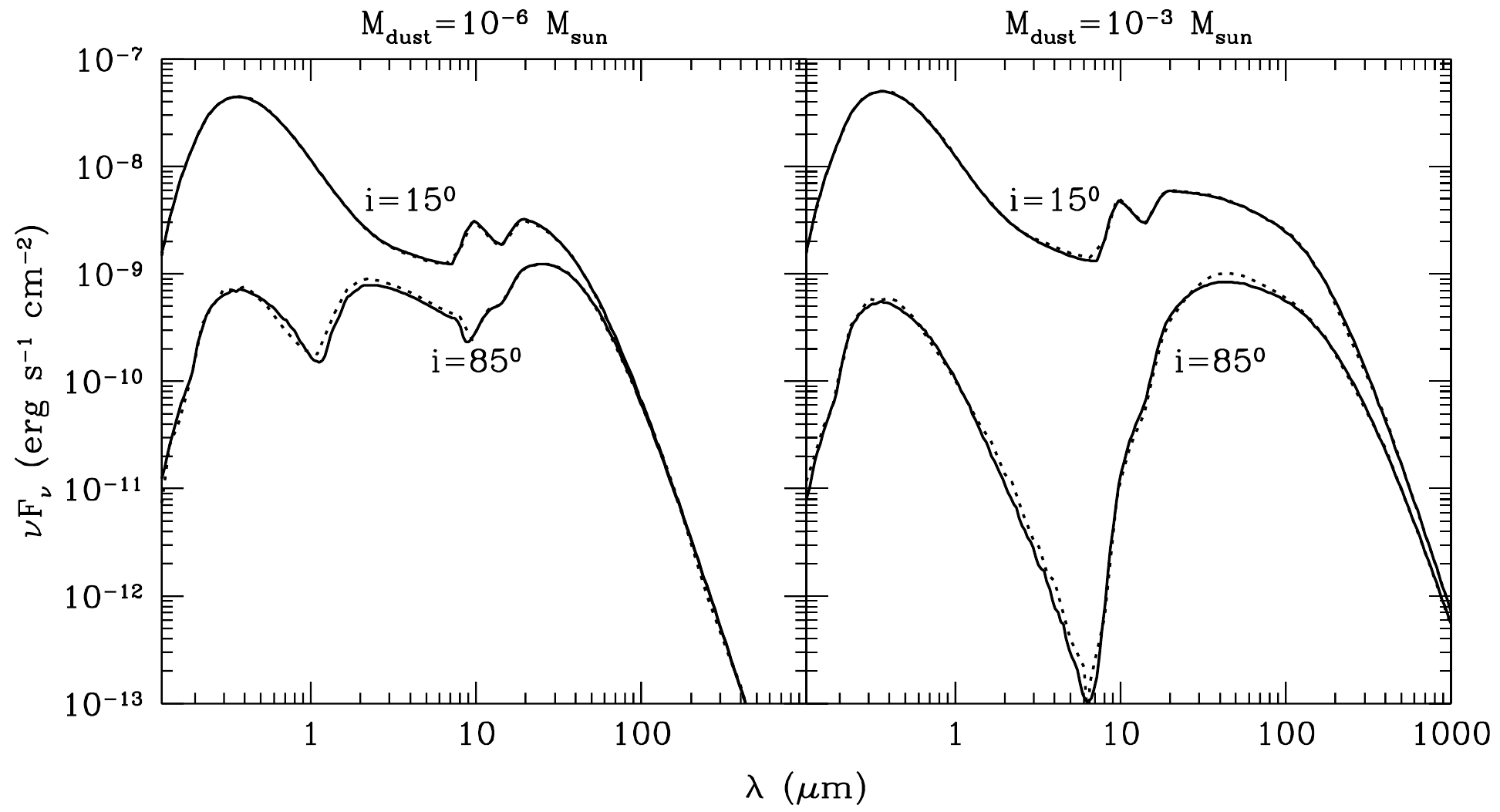}}}
\caption{The resulting SEDs as computed for the two disk masses. The SEDs are shown for two disk inclination angles. The solid line gives the resulting spectrum from MCMax while the dotted line shows the result as computed by RADMC. The distance to the object is 150\,pc.}
\label{fig:results100M}
\end{figure*}

The comparison of the computed temperature structures is shown in Fig.~\ref{fig:compare codes}. As the error we plot
\begin{equation}
\frac{T_\mathrm{RADMC}-T_\mathrm{MCMax}}{T_\mathrm{RADMC}+T_\mathrm{MCMax}}.
\end{equation}
For the model with $10^{-6}\,M_{\sun}$ of dust the differences are below 6\% everywhere in the disk. All differences seem to be caused by interpolating between the different spatial grids used in both codes. For the model with $10^{-3}\,M_{\sun}$ of dust, the differences are slightly larger. In this case we can distinguish three types of differences:
\begin{enumerate}
\item Those caused by the interpolation of the spatial grids.
\item Those caused by the fact that the inner boundary in the RADMC code is smoothed, while in the MCMax code it is sharp.
\item Errors caused by photon noise and numerical implementation of the methods. 
\end{enumerate}
At places in the disk where the temperature gradient is large the errors due to interpolation are most important. In these regions, these errors can get up to $\sim6\%$. Differences 2) only play a role at the innermost edge of the disk. Here we find errors, which are most likely caused by this effect, up to $\sim30\%$. These errors occur only in an extremely small region of the disk, i.e. the inner $2\cdot10^{-4}\,$AU. Note that the temperature of the emitting inner rim is the same for both codes, as is the temperature gradient expressed in \emph{optical depths} (which is important for determining the inner rim emission). The smoothing of the inner rim as used in RADMC, causes a slightly different temperature gradient as a function of \emph{radius} since the optical depth gradient is different. The differences occur only inside $5\cdot10^{-5}\,$AU from the inner rim of the dust disk. Errors that cannot be attributed to effects 1 or 2 are due to numerical implementation of the equations and statistical noise. These can be considered the true differences caused by the differences in the codes. They are typically small, except for the midplane region. In this region, the photon count is low and the exact implementation of the method becomes very important. The true differences are typically $\sim15\%$ close to the midplane and smaller than a few percent everywhere else in the disk. We thus conclude that overall the two codes agree very well on the computed temperature structures.

The resulting spectral energy distributions for a disk at a distance of 150\,pc at two different inclinations are shown in Fig.~\ref{fig:results100M}. The differences between the RADMC and MCMax models are small, although for the high inclination cases differences can be observed even on a log-log scale. Since the shape of the spectra is identical, the most likely cause for this is the spatial grid chosen by both codes. 

We conclude that the accuracy of the codes is very good, and the results compare very well.

\end{document}